# Testing Parametric Distribution Family Assumptions via Differences in Differential Entropy

Ron Mittelhammer,[1] George Judge[2] and Miguel Henry[3]

**ABSTRACT**

We introduce a broadly applicable statistical procedure for testing which parametric distribution family generated a random sample of data. The method, termed the Difference in Differential Entropy (DDE) test, provides a unified framework applicable to a wide range of distributional families, with asymptotic validity grounded in established maximum likelihood, bootstrap, and kernel density estimation principles. The test is straightforward to implement, computationally efficient, and requires no user-defined tuning parameters or complex specialized regularity conditions. It compares an MLE-based estimate of differential entropy under the null hypothesis with a nonparametric bootstrapped kernel density estimate, using their divergence as an information-theoretic measure of model fit. The test procedure is constructive in the sense of being informative regardless of whether the null hypothesis is rejected or not, where in the latter case the outcome suggests that the hypothesized distribution can be close to the actual distribution of the data in shape and probability implications. Monte Carlo experiments demonstrate its notable size accuracy and power even in relatively small samples, and three empirical applications using classical datasets from distinct domains illustrate the method's practical utility.

[1] Regents Professor of Economic Sciences, School of Economic Sciences, Room 121C Hulbert Hall, Washington State University, Pullman, WA (email: mittelha@wsu.edu). Corresponding author.
[2] Professor in the Graduate School, University of California, Berkeley, CA (email: gjudge@berkeley.edu).
[3] Economist, OnPoint Analytics, Emeryville, CA (email: mhenry@onpointanalytics.com).

## 1. Introduction

Testing assumptions about parametric probability distributions remains an active research area in statistics and econometrics. Although this literature has expanded substantially since Pearson (1900), unified and broadly applicable omnibus methods with significant power for testing a wide range of distributional hypotheses remain elusive, despite several important developments (e.g., Bowman and Shenton 1975; Epps and Pulley 1983; Zoubir and Arnold 1996; Doornik and Hansen 2008; Meintanis 2011; and Wyłomańska et al. 2020). The existing literature offers a variety of methods for specific parametric distribution families, which often involve complex and idiosyncratic regularity conditions, can result in non-standard probability distributions under both null and alternative hypotheses, and/or require specifying tuning parameters. These methods also often impose restrictive implementation requirements that limit their scope of application.

In empirical economics, correctly specifying the distributional family of random variables is crucial for valid statistical inference, risk assessment, and forecasting. Common applications include such things as error terms in econometric models, modeling asset returns in finance, duration data in labor economics, and income distributions in public economics. Traditional goodness-of-fit tests like Kolmogorov-Smirnov or Anderson-Darling are not dominant in power against all alternatives and do not straightforwardly accommodate composite hypotheses in which parameters are unknown. Zhang (2023) underscores the growing relevance of entropy-based approaches in modern data science and motivates renewed attention to their theoretical foundations.

This paper contributes an application of information theoretic statistical testing methodology that assesses the validity of composite hypotheses regarding the parametric distributional family that is postulated to govern the outcomes observed in a random sample of data. We utilize the concept of differential entropy, as well as maximum entropy distributions where applicable. The primary objective is to provide a unified testing methodology that is flexible, straightforwardly implementable, and valid for testing a broad range of parametric distribution families using a sample of data. We focus on continuous random variables and their associated parametric distribution families, although the method can be extended to discrete random variables and distributions using similar information theoretic principles, *mutatis mutandis*. A key advantage of this approach is the application of the same consistent and straightforward testing framework, including the same test statistic and decision rules, for assessing hypotheses about a very wide range of parametric distribution families commonly used in statistical and econometric practice. Moreover, the null distribution used for establishing critical values for tests, as well probability values associated with a null hypothesis, are self-generating and do not rely on fixed asymptotic or limiting distributions.

The paper is structured as follows. Section 2 provides a review of literature. Section 3 introduces foundational background on differential entropy, the general motivation for a test based on its value, and the concept of maximum entropy distribution. Section 4 presents the methodological details underlying parametric and nonparametric estimation of differential entropy for random samples of data. Section 5 defines the difference in differential entropy (DDE) testing approach for evaluating null hypotheses about parametric families of distributions and discusses its sampling properties. Section 6 reports Monte Carlo simulations assessing the size and power of the DDE test. Section 7 applies the methodology to three classic datasets drawn from the historical literature, and section 8 offers a summary and concluding remarks.

## 2 Literature Review

### *2.1 Historical Development of Distribution Testing*

Since Pearson's (1900) seminal work introducing the chi-square goodness-of-fit test and its variants (e.g., the Cramér–von Mises and Kolmogorov–Smirnov tests), numerous procedures have been proposed for testing distributional functional forms. D'Agostino and Stephens (1986) provide a comprehensive survey of early work, and additional historical reviews include Thode's (2002) examination of normality tests and Shapiro and Brain's (1982) extensive coverage of various distributional assumptions relating to tests from earlier years. The econometrics literature features notable contributions from Lee, who developed tests for distributional assumptions in stochastic frontier functions (1983), univariate normal distributions (with Bera and Jarque 1984), bivariate normal distributions (1984), and count data models (1986).

More recent developments in the econometrics literature include Stengos and Wu's (2009) information-theoretic distribution test with application to normality, Bontemps and Meddahi's (2012) test based on moment conditions, Bera *et al.*'s (2016) normality test using the quantile-mean covariance function, Amengual et al.'s (2020) goodness-of-fit tests for parametric distributions employing regularization methods to evaluate differences between theoretical and empirical characteristic functions, and Mittelhammer et al.'s (2022) entropy-based approach for nonparametrically testing simple hypotheses relating to specific probability distributions postulated to have generated an observed data sample.

### *2.2 Nonparametric Testing Approaches*

One of the earliest nonparametric tests of a distributional hypothesis is the Wald and Wolfowitz (1940) independence test (WW test), which is based on the concept of runs in a sequence of sample observations arranged in increasing order of magnitude. Although the WW test has a familiar chi-square asymptotic distribution, it lacks an explicit provision for assessing the identical distribution assumption, which is typically treated as a maintained hypothesis in empirical applications (see, e.g., Fama 1965). Several variations of runs-type tests have emerged, including Goodman's (1958) simplified runs test and Cho and White's (2011) generalized runs test. Other nonparametric approaches like the Mann–Kendall test (Mann 1945; Kendall 1975) and Bartels' (1982) rank-based test have been used to detect trends and specific departures from *iid* behavior.

Despite these developments, no single nonparametric testing approach has proven to dominate in the sense of being universally appropriate and powerful across all sampling situations – a fundamentally unlikely prospect given, among other considerations, the vast array of possible alternative population distributions. However, pursuing the objectives of wide applicability and the use of a uniformly consistent testing approach, the DDE presented ahead provides an advance in that direction, while incorporating a nonparametric estimation element as one of its definitional components.

### *2.3 Entropy-based Testing Methods*

The development of entropy-based tests began with pioneering contributions that highlighted their favorable performance relative to alternative approaches. Vasicek (1976) developed a consistent entropy-based test for the composite hypothesis of normality. Building on this line of research, Zhu et al. (1995) developed an entropy-based approach for testing multivariable normality, while Dudewicz and van der Meulen (1981) proposed a similar approach for testing uniformity. The Vasicek (1976) test was later adapted by Lee (2016) for use in testing some composite hypotheses under suitable regularity conditions, and the resulting test was shown to be asymptotically normally distributed under the stated conditions.

Recent studies by Matilla-García and Marín (2008) and Canovas and Guillamon (2009) assess the *iid* null hypothesis using permutation entropy (PE). Their approach requires a data series with a natural time-dependent order – a common albeit not a universal feature of econometric applications. The PE method offers some advantages in that it does not require rigid model-based assumptions and remains invariant to monotonic data transformations. However, its explicit design for time-series observations on scalar variables does not provide for extending the nonparametric independence test to multivariate contexts, and the testing methodology does not assess parametric families of distribution per se.

Hong and White (2005) developed an alternative test of serial independence for scalar time series based on standard entropy concepts. Their approach, which is asymptotically locally more powerful than Robinson's (1991) smoothed nonparametric modified entropy measure, relies on Kullback–Leibler divergence (Kullback and Leibler 1951) and the fundamental principle that a joint density of serial observations factors into the product of its marginal distributions if and only if the observations are independent. However, both the Hong–White and Robinson tests exhibit a lack of local power, and the issue becomes more pronounced as the dimensionality of the data increases (Ai et al. 2024). Another issue in these implementations is their exclusive focus on kernel density estimation methods, and choices of kernels and bandwidth parameters that have the potential to lead to differing test outcomes. Racine and Maasoumi (2007; at p. 559) note that the sensitivity of asymptotic entropy-based tests to bandwidth choice is "partly because the bandwidth vanishes asymptotically," so that the "asymptotic-based null distribution would not depend on the bandwidth, while the value of the test statistic does so directly." Furthermore, the finite sample level of these tests may diverge from asymptotic levels. Hong and White (2005, at p. 850) acknowledge that asymptotic theory may not work well even with relatively large samples when applying these tests.

No current entropy-based testing framework has been demonstrated to be uniformly superior, and limitations include such things as restrictions in applicability, idiosyncratic and complex regularity conditions required to ensure asymptotic validity, and rates at which accurate test size and notable test power is achieved in finite samples. These and other issues relating to entropy-based testing procedures are described in Matilla-García and Marín (2008) as well as in Cadirci et al.'s (2022) recent contribution in providing an entropy-based test for generalized Gaussian distributions. The development of improved testing procedures remains an ongoing and relevant research objective. Addressing some of these methodological deficiencies, Mittelhammer et al. (2022) introduced a flexible, unified, and widely applicable nonparametric entropy test methodology for assessing the validity of simple hypotheses about specific members of a parametric family, with consistent asymptotic chi-square calibrations for the test statistic under straightforward regularity conditions. This paper provides a more general approach that is applicable for testing simple as well as composite hypotheses postulating entire parametric families of distributions thought to be consistent with the data generating process underlying an observed sample of data, and do not rely on asymptotic chi-square calibrations for defining critical values or probability values of the test.

## 3. Differential Entropy

### *3.1 Basic Definition*

The foundational concept underlying the statistical testing procedure is differential entropy (DE), originally introduced by Shannon (1948) and later formalized by Kolmogorov (1956). Differential entropy provides a fundamental measure of distributional uncertainty. For a continuous random variable $\boldsymbol{X}$ with probability distribution $f(x;\boldsymbol{\theta})$, differential entropy is defined as

$$DE(f;\boldsymbol{\theta}) = -\int_x f(x;\boldsymbol{\theta}) \ln\left(f(x;\boldsymbol{\theta})\right) dx \tag{1}$$

### *3.2 Testing Principle Based on DE*

Our test statistic exploits a fundamental insight: both parametric Maximum Likelihood Estimator (MLE) and nonparametric Kernel Density Estimator (KDE) estimates of differential entropy converge to each other when the assumed parametric distribution family $f(x;\boldsymbol{\theta}), \boldsymbol{\theta} \in \Omega$ underlying the data generating process is correct, but they diverge otherwise. This encompassing principle can be used to assess a wide range of commonly assumed population distributions implemented in empirical practice and can be applied to both simple and composite specifications of distributional hypotheses, i.e.,

$$H_0 : X_i\text{'}s \sim iid\ f(x;\boldsymbol{\theta})\ \ for\ \ \boldsymbol{\theta} = \boldsymbol{\theta}_0\ \ or\ \ \boldsymbol{\theta} \in \Omega \tag{2}$$

where the random sample under consideration is $\mathbf{X} = \{X_1, X_2, \cdots, X_n\}$.

### *3.3 Maximum Entropy Distributions (MEDs)*

A useful insight that contributes to the theoretical foundation of the differential entropy testing methodology is that many parametric distributions commonly used in statistical and econometric applications are Maximum Entropy Distributions – they maximize differential entropy subject to moment constraints and random variable supports. A general characterization of a MED is given by

$$f_{MaxEnt}\left(x;\boldsymbol{\theta}\right)=\arg\max_{f}\left\{-\int_{x} f\left(x;\boldsymbol{\theta}\right)\ln\left(f\left(x;\boldsymbol{\theta}\right)\right)dx \quad s.t. \quad E\left[\mathbf{g}\left(X;\boldsymbol{\theta}\right)=\mathbf{0}\right]\wedge x\in R\left(X\right)\right\} \tag{3}$$

where $\mathbf{g}\left(X;\boldsymbol{\theta}\right)=\mathbf{0}$ are constraint functions and $R\left(X\right)$ denotes the range or support of $X$.

As an explicit example of a MED arising from a solution to (3), consider the following maximum differential entropy problem:

$$\arg\max_{f}\left\{-\int_{x} f\left(x;u,\sigma^2\right)\ln\left(f\left(x;u,\sigma^2\right)\right)dx \quad s.t. \quad E\begin{bmatrix} X \\ X^2 \end{bmatrix}=\begin{bmatrix} u \\ u^2+\sigma^2 \end{bmatrix}\wedge x\in\mathbb{R}\right\} \tag{4}$$

The solution is the Normal distribution

$$f_{MaxEnt}\left(x;\boldsymbol{\theta}\right)=\frac{1}{\sqrt{2\pi\sigma^2}}\exp\left(-\frac{1}{2}\left(\frac{x-u}{\sigma}\right)^2\right) for\ \boldsymbol{\theta}=\begin{Bmatrix} u \\ \sigma^2 \end{Bmatrix}\in\Omega=\mathbb{R}\times\mathbb{R}_+\wedge x\in\mathbb{R} \tag{5}$$

with $\mathbb{R}=\left(-\infty,\infty\right)$, $\mathbb{R}_+=\left(0,\infty\right)$, and the DE value is defined by

$$DE\left(N\left(x;\,u,\sigma^2\right)\right)=\frac{1}{2}\ln\left(2\pi e\sigma^2\right) \tag{6}$$

Appendix A provides MED solutions and associated values of differential entropy for a wide variety of parametric families of distributions utilized in empirical practice.

## 4. Parametric and Nonparametric Differential Entropy Estimation

The DDE testing methodology employs two different estimators of DE. The first is a parametric MLE for DE based on the hypothesized null distribution functional form. The second is a KDE, which provides a nonparametric representation of the probability density function component of the integrand in (1) used to define the DE. The functional details and some key interpretations of each of these estimators are presented in this section.

*4.1 MLE of Differential Entropy*

The MLE of differential entropy is derived using maximum likelihood estimates[4] $\hat{\boldsymbol{\theta}}_{ML}$ of the parameters in the parametric family of distributions specified under the null hypothesis $H_0$. While the testing methodology is generally designed to test composite hypotheses –that is, entire parametric families of distributions – the test can be readily adapted to test a simple null hypothesis by replacing the MLE with $\hat{\boldsymbol{\theta}}_{ML} = \boldsymbol{\theta}_0$. The MLE of differential entropy is defined by

$$DE_{ML}(f) = -\int_x f\left(x;\hat{\boldsymbol{\theta}}_{ML}\right)\ln\left(f\left(x;\hat{\boldsymbol{\theta}}_{ML}\right)\right)dx \tag{7}$$

via the maximum likelihood invariance principle.

The maximum likelihood value for estimating the null-hypothesized component of DDE, $DE_{ML}(f)$, is obtained in the usual way based on the solution to

$$\hat{\boldsymbol{\theta}}_{ML} = \underset{\boldsymbol{\theta}\in\Omega}{\arg\max}\left(n^{-1}\sum_{i=1}^{n}\ln\left(f\left(x_i;\boldsymbol{\theta}\right)\right)\right). \tag{8}$$

Then $\hat{\boldsymbol{\theta}}_{ML}$ is substituted into (7) to obtain the value of $DE_{ML}(f)$. If the distribution under consideration is a maximum entropy (MaxEnt) distribution the integral in (7) will then generally admit a convenient closed-form analytical functional expression, as noted in Section 3 (See Appendix A for a collection of other MEDs). In such cases, the value of $DE_{ML}(f)$ can be obtained using a straightforward plug-in method as $DE_{ML}(f) = MaxEnt\left(\hat{\boldsymbol{\theta}}_{ML}\right)$. For example, suppose the null hypothesis is defined as

$$H_0 : x_i \sim iid\ N\left(x; u, \sigma^2\right),\ i = 1,2,\cdots,n. \tag{9}$$

Then the MLE of differential entropy is estimating the maximum differential entropy,

[4] In situations where obtaining the MLE estimate is challenging or intractable, it is possible to substitute alternative consistent estimators of the parameters, e.g., method of moment estimators, which will allow the testing framework to proceed with properties similar to that of the DDE test based on MLEs.

$$DE_{ML}\left(N\left(x;\, u,\sigma^2\right)\right)=\frac{1}{2}\ln\left(2\pi e\hat{\sigma}^2\right) \tag{10}$$

where $\hat{\sigma}_{ML}^2 = n^{-1}\sum_{i=1}^{n}\left(x_i-\bar{x}\right)^2$ .

4.1.1 *MLE Differential Entropy Bias*

In formulating any testing procedure that utilizes the MLE of differential entropy, it is important to recognize that differential entropy is typically a nonlinear function of the MLE of the parameters, $\hat{\boldsymbol{\theta}}$. Consequently, even if $\hat{\boldsymbol{\theta}}$ is itself unbiased, the corresponding estimator of differential entropy will generally exhibit some degree of small sample bias. The explicit functional form of this bias depends on the null distribution specified and its associated parametric functional form. More generally, under standard ML regularity conditions, the bias of the MLE of differential entropy is given by (See Appendix B, Theorem B.1):

$$Bias_{MLE}(n)=\frac{1}{2n}tr\left(I^{-1}\left(\boldsymbol{\theta}_0\right)\nabla_{\boldsymbol{\theta}}^2 H\left(\boldsymbol{\theta}_0\right)\right)+o\left(n^{-1}\right) \tag{11}$$

where $I\left(\boldsymbol{\theta}_0\right)$ is Fisher's information matrix, and $\nabla_{\boldsymbol{\theta}}^2 H\left(\boldsymbol{\theta}_0\right)$ is the Hessian of the differential entropy functional.

Given the bias representation in (11), an empirical first order correction to the MLE of differential entropy could be implemented in principle by subtracting an estimate of the leading term in (11) from the MLE estimate of differential entropy itself, based on the MLE $\hat{\boldsymbol{\theta}}_{ML}$. A standard concern with such plug-in estimators is that it may introduce an additional source of sampling error in generating the estimate of differential entropy. Whether this becomes corrective or problematic depends on the parametric family of distributions under consideration. Complete solutions based on (11) for the functional forms of the first order bias corrections corresponding to four families of distributions used in Section 6 to simulate the size and power of the test procedure are provided in Table 1. For the normal, exponential, and LaPlace parametric families, the first-order bias correction does not depend on any unknown parameters and is therefore first-order exact. In contrast, for the more general gamma parametric family of distributions the bias correction depends on an unknown distributional parameter, which must be estimated from the data to obtain a plug-in estimate of the bias.

**Table 1. Estimation Bias and Correction for $DE_{ML}(f)$**

| Distribution Family | $\mathbf{Bias}_{ML}^{f}(n)$ | Unbiased Estimator ($0(n^{-1})$) of $DE_{ML}(f)$: $DE_{ML}(f)-\mathbf{Bias}_{ML}^{f}(n)$ |
|---|---|---|
| $N(u,\sigma^2)$ | $\frac{1}{2}\left[\psi\left(\frac{n-1}{2}\right)-\ln\left(\frac{n}{2}\right)\right]$ | $\frac{1}{2}\left(\ln\left(2\pi e\hat{\sigma}^2\right)-\left[\psi\left(\frac{n-1}{2}\right)-\ln\left(\frac{n}{2}\right)\right]\right)$ |
| $Exponential(\theta)$ | $\frac{1}{2n}$ | $1+\ln\left(\hat{\theta}\right)-\frac{1}{2n}$ |
| $Gamma(\alpha,\beta)$ | $\frac{1}{2n\alpha}+\left(\frac{1-\alpha}{2n}\right)\left[1-(\alpha-1)\psi'(\alpha)\right]$ | $\ln\left(\hat{\beta}\Gamma(\hat{\alpha})\right)+(1-\hat{\alpha})\psi(\hat{\alpha})+\hat{\alpha}$ $-\left[\frac{1}{2n\hat{\alpha}}+\left(\frac{1-\hat{\alpha}}{2n}\right)\left[1-(\hat{\alpha}-1)\psi'(\hat{\alpha})\right]\right]$ |
| $LaPlace(u,b)$ | $-\frac{1}{2n}$ | $1+\ln\left(2\hat{b}\right)+\frac{1}{2n}$ |

**Note:** $\psi(\bullet)$ is the digamma function, $\psi'$ is the trigamma function, and all parameter estimators refer to MLEs.

### 4.2 *KDE of Differential Entropy*

The KDE of differential entropy is defined by

$$DE_{KDE}\left(\hat{f};h\right)=-\int_{x}\hat{f}(x;h)\ln\left(\hat{f}(x;h)\right)dx \tag{12}$$

where $\hat{f}(x;h)$ is a KDE applied to the sample data $\mathbf{x}$ based on a bandwidth $h$. The implementation in this paper utilizes a Gaussian kernel to define the KDE, as

$$\hat{f}(x;h)=\frac{1}{nh}\sum_{i=1}^{n}K\left(\frac{x-x_i}{h}\right)\ \ where\ K(u)\equiv\frac{1}{\sqrt{2\pi}}\exp\left(-\frac{u^2}{2}\right). \tag{13}$$

The value of the integral in (12) can be calculated via numerical integration procedures such as the adaptive quadrature approach. In our applications ahead, we implement that numerical integration approach as implemented in GAUSS (version 25) through the *integrate1d* procedure with default convergence settings. The integration was stable and fast in all applications, as expected given the differentiability and smoothness properties of (13) and the integrand in (12). Integration ranges for the entropy functional were determined using quantile-based intervals on the appropriate numerical scale (log scale on $\mathbb{R}_+$, raw scale on $\mathbb{R}$), where upper and lower limits of integration were based on .999 and .001 quantiles of the data, adjusted up or down by multiples of the bandwidth to enhance numerical stability and prevent tail over-smoothing. The resulting value of $DE_{KDE}\left(\hat{f};h\right)$ provides a fully data-based, nonparametric empirical estimate of the differential entropy for the given sample of data.

Several alternative well-known kernels have been used to define KDE estimates, including the Expanechnikov, Tri-weight, and Uniform kernels. However, the Gaussian kernel is generally regarded as well-suited – and in some respects optimal – for specific use in entropy estimation (Hall and Morton (1993); Giné and Nickl (2008); and Berrett et al. (2019)). Its $C^{\infty}$ smoothness mitigates instability when evaluating $\ln\left(f\right)$ and ensures continuity when integrating functionals such as entropy. Moreover, unlike some other kernels, the Gaussian kernel is strictly positive-valued everywhere, thereby avoiding the problem of $\ln\left(f\right)$ diverging at zero-values and guaranteeing that it is always well-defined and finite. In addition, the Gaussian kernel's tails decay at a faster rate than any polynomial, which reduces truncation error in numerical integration and lowers variance when the underlying sampling density has light-to-moderately heavy tails. Finally, in terms of integrated squared error (ISE) for functionals such as differential entropy, the Gaussian kernel has been found to perform favorably in practice, owing to its rapid tail decay and smoothness properties (Hall and Morton 1993; Beirlant et al. 1997; and Joe 1989).

#### *4.2.1 KDE Differential Entropy Estimation When Supports are Nonnegative*

A particular bias problem in KDE-based estimation of differential entropy arises when the method is applied to null distributions with nonnegative support, or more generally to distributions with finite lower and/or upper bounds, while using a Gaussian kernel. Specifically, evaluations of the kernel near boundaries of the support can distort the density and differential entropy estimates by including values of $x$ that are not observable under the null. In these situations, an effective approach is to apply KDE estimation to the density of $ln\left(x\right)$, and then express the original KDE estimate of differential entropy as a function of the density of $ln\left(x\right)$, supplemented by an adjustment term involving the sample mean of the $ln\left(x\right)$ outcomes. The specific representation is given by

$$DE_{KDE}\left(\hat{f};h\right)=-\int_{0}^{\infty}\hat{f}\left(x;h\right)\ln\left(\hat{f}\left(x;h\right)\right)dx=-\int_{-\infty}^{\infty}\hat{g}\left(y;h\right)\ln\left(\hat{g}\left(y;h\right)\right)dy+\bar{y},\ where\ y=\ln\left(x\right) \quad (14)$$

Estimator (14) represents the KDE estimate of differential entropy when the Kernel estimation procedure is applied in *ln*-space. The derivation of this result is provided in Theorem B.2 of Appendix B. Key references supporting the application of kernel density estimation in *ln*-space, particularly for mitigating boundary bias, include Jones et al. (1995), Wand et al. (1991), and Marron and Ruppert (1994). In our analysis, we implement the *ln*-space approach to any null distributions having nonnegative-valued support.

#### *4.2.2 KDE Differential Entropy Bias*

In addition to the support boundary bias issue discussed above, it is known that the KDE approach exhibits varying degrees of small-sample bias when used to estimate differential entropy (12). For Gaussian kernels (or more generally, symmetric kernels), the form of the leading order bias term takes the following form (See Joe 1989; Beirlant et al. 1997; Hall and Morton 1993):

$$Bias_{KDE}\left(\hat{f};h\right)=\frac{h^2}{4}\int f\left(x\right)\left[\frac{\partial^2 f\left(x\right)/\partial x^2}{f\left(x\right)}-\left(\frac{\partial f\left(x\right)/\partial x}{f\left(x\right)}\right)^2\right]dx+\frac{\int K^2\left(u\right)du}{2nh}. \quad (15)$$

The first term in the sum is typically referred to as the *smoothing bias*, while the second term is the *variance bias*. In principle, KDE differential entropy bias can be mitigated by subtraction, i.e., $DE_{KDE}\left(\hat{f};h\right)-Bias_{KDE}\left(\hat{f};h\right)$. However, the smoothing component of the bias adjustment depends on the functional form of the underlying probability density $f\left(x\right)$ and generally requires an integration operation, which may need to be implemented numerically. A more direct approach to calculating $Bias_{KDE}\left(\hat{f};h\right)$ is available if $f\left(x\right)$ belongs to certain parametric families of distributions (e.g., normal, exponential, gamma, among others), for which relatively straightforward closed-form solutions exist for (15). For example, in the case of $N\left(u,\sigma^2\right)$, it can be shown that $Bias_{KDE}\left(\hat{f};h\right)=-\frac{h^2}{4\sigma^2}+\left(4nh\sqrt{\pi}\right)^{-1}$, regardless of the values of $u$. This expression represents the sum of the smoothing and variance biases. Moreover, when a symmetric non-Gaussian kernel is used, the order of the variance bias adjustment $\left(nh\right)^{-1}$ still applies, but the constant multiplier $\left(4\sqrt{\pi}\right)^{-1}$ will be replaced by a kernel-dependent multiplier. This scenario holds under Silverman's classic bandwidth choice $1.06n^{-1/5}\hat{\sigma}_{\mathbf{X}}$ (Joe 1989; Beirlant et al. 1997; Hall and Morton 1993; Moon et al. 1995). Additional closed-form solutions to (15) are available depending on the hypothesized null distribution, and solutions for three null distributions simulated in Section 6 are provided in Table 2.

As with the earlier discussion of bias in the MLE, implementing a first-order bias correction by subtracting an estimated value of the bias from the KDE of differential entropy introduces additional variation, since unknown quantities in these expressions must themselves be estimated. Thus, implementing such a bias correction cannot be unequivocally claimed to provide a uniform improvement in the resulting estimate of differential entropy.

**Table 2. Smoothing Bias Examples and Associated Unbiased Estimators**

| **Distribution Family** | $\frac{h^2}{4}\int f(x)\left[\frac{\partial^2 f(x)/\partial x^2}{f(x)} - \left(\frac{\partial f(x)/\partial x}{f(x)}\right)^2\right]dx$ | **Unbiased Estimator** $\leq O\left(n^{-1}\right)$ |
|---|---|---|
| $N\left(u,\sigma^2\right)$ | $-\frac{h^2}{4\sigma^2}$ | $-\frac{h^2}{4\hat{\sigma}^2}\left(\frac{n-3}{n-1}\right)$ |
| $Ln\,Exponential\left(\theta\right)$ | $-\frac{h^2}{4}$ | $-\frac{h^2}{4}$ |
| $Ln\,Gamma\left(\alpha,\beta\right)$ | $-\frac{h^2}{4}\alpha$ | $-\frac{h^2}{4}\left(\hat{\alpha}-\left(2n\right)^{-1}\left(1-\hat{\alpha}\right)\right)$ |
| $LaPlace\left(u,b\right)$ | $-\frac{h^2}{4b^2}$ | $-\frac{h^2}{2}\left(\frac{1-5n^{-1}}{s^2}\right)$ |

**Note:** The parameter estimators $\hat{\sigma}^2$ *and* $\hat{\alpha}$ are MLEs, and $s^2$ is the unbiased sample variance estimator. In the case of the LaPlace distribution, the existence of the cusp in the density violates the regularity class (twice continuous differentiability) assumed for the smoothing bias expression, and so the result is extended by implementing generalized derivatives. Regarding bias corrections, also see Cordeiro and McCullagh (1991).

#### *4.2.3 KDE Bandwidth Choice*

The definition for the bandwidth $h$ of the KDE is of the familiar form

$$h = k\left(n\right)c\,\hat{\sigma}\,n^{-1/5} \tag{16}$$

where $n$ is the sample size, $\hat{\sigma}$ is an estimate of scale appropriate to the data used in the KDE,

which in our applications is either the original data scale for $x \in \mathbb{R}$ or log-transformed data for $x \in \mathbb{R}_+$, $c$ is a shape-dependent multiplicative adjustment factor, and $k(n)$ is a small sample inflation factor that is defined ahead. The rate $n^{-1/5}$ follows the classic mean-squared-error (MSE) optimal rate for one-dimensional KDE plug-in estimators (Hall and Marron 1987; Joe 1989; Wand and Jones 1995; Beirlant et al. 1997). This rate also minimizes the combined bias $O(h^2)$ and variance $O((nh)^{-1})$ of functionals of the KDE, including the entropy functional $H(f) = -\int f(x)\ln\left(f(x)\right)dx$. Although functionals of a density can sometimes justify different smoothing rates, the plug-in entropy estimator $-\int \hat{f}_h(x)\ln\left(\hat{f}_h(x)\right)dx$ shares the same local bias and variance expansion as the integrated squared error (ISE) for density estimation. Hence, the optimal asymptotic rate for minimizing MSE of estimated entropy remains $h \propto n^{-1/5}$ (Joe 1989; Beirlant et al. 1997, Section 3.3; Wand and Jones 1995, Ch. 3).

Entropy estimation is sensitive to smoothing bias, e.g., inflated bandwidths can flatten modes and extend tails, leading to systematic overestimation of entropy. The rate multiplier $c$ plays a role in mitigating bias. It is specified here based on theoretical and empirical recommendations found in the KDE literature applied to entropy estimation, and it is adaptive to the shape characteristics of various null distributions. In the interest of providing an automated definition of bandwidth applicable to a wide scope of null distributions, we focused attention on four broad null distribution categories arising in practice: Gaussian, "near-Gaussian" on $\mathbb{R}$, non-Gaussian unimodal distributions on $\mathbb{R}$, and unimodal right-skewed distributions on $\mathbb{R}_+$. Table 3 summarizes the definitions of $c$ based on the category of null distribution to which the entropy estimator is applied. Detailed literature and simulation-based rationale for these definitions is given in Appendix C.

**Table 3. Shape-Based Definitions for $h = k(n)\, c\, \hat{\sigma}\, n^{-1/5}$**

| **Null** | $c$ | **Conditions** | $k(n)$ | $\hat{\sigma}$ Scale |
|---|---|---|---|---|
| Gaussian | 1 | NA | 1 | $X$ |
| Near-Gaussian | | $\hat{\kappa} \in [2.0,\ 4.0]$ <br> $x \in \mathbb{R}, \lvert \hat{s}_{skew} \rvert \le .5$ | | $X$ |
| Non-Gaussian | $1 + 0.1 \log_2 \left( \hat{\gamma}_{\text{KURT}} \right)$ | $\hat{\kappa} \notin [2.0,\ 4.0]$ <br> $x \in \mathbb{R}$ <br> $\lvert \hat{s}_{skew} \rvert > .5$ | $\begin{Bmatrix} 1.25 - 0.25\left(\frac{n-50}{50}\right) \\ 1 \end{Bmatrix} \textit{for} \begin{Bmatrix} n < 100 \\ n \ge 100 \end{Bmatrix}$ | $X$ |
| Right-Skewed Unimodal | | $x \in \mathbb{R}_+$ | | $ln(X)$ |

**Notes:** $\hat{\gamma}_{\text{kurt}} = \frac{\kappa_0}{\tau(\hat{\kappa})}$, $\tau(\hat{\kappa}) = \min\{\max(\hat{\kappa}, 2), 10\}$, $\hat{\kappa} = \frac{\frac{1}{n}\sum_{i=1}^{n}(X_i - \bar{X})^4}{\left(\frac{1}{n}\sum_{i=1}^{n}(X_i - \bar{X})^2\right)^2}$, $\kappa_0$ is the kurtosis of the null distribution (e.g., 3 for Gaussian, $3 + 6/\hat{\alpha}$ for Gamma), $\hat{s}_{skew} = \frac{\frac{1}{n}\sum_{i=1}^{n}(X_i - \bar{X})^3}{\left(\frac{1}{n}\sum_{i=1}^{n}(X_i - \bar{X})^2\right)^{3/2}}$, and $c$ is constrained to the interval $[.85,\ 1.15]$.

## 5. The Difference in Differential Entropy (DDE) Test

The statistical test of the parametric family of distributions hypothesis (2) is based on the estimated differential entropy difference between the MLE and the KDE of differential entropy

$$DDE(f) = DE_{ML}(f) - DE_{KDE}\left(\hat{f}; h\right). \tag{17}$$

Under the null hypothesis, and given the typical regularity conditions supporting the asymptotic behavior of ML estimators, and using bandwidth specifications with the rate indicated in (16), both estimators of differential entropy converge to the true value of differential entropy, and their divergence is of order $O_p\left(n^{-2/5}\right)$, as

$$DE_{ML}(f) \wedge DE_{KDE}\left(\hat{f}; h\right) \xrightarrow{p} DE(f) \text{ and } DE_{ML}(f) - DE_{KDE}\left(\hat{f}; h\right) = O_p\left(n^{-2/5}\right). \tag{18}$$

However, if the hypothesized family of probability distributions defined by $H_0$ does not encompass the true probability distribution underlying the outcomes of the random sample of data, then the DDE estimates diverges from zero and is indicative of a rejection of the null hypothesis as well as misspecification of the probability model assumed for the data generating process. The test outcome is determined by whether the DDE value is statistically significant. Motivation for both the convergence of the two estimators under $H_0$ and the divergence when the null hypothesis is false is provided in Appendix B, along with motivation for the asymptotic normality of the test statistic.

### *5.1 Intuition Underlying the DDE Testing Approach*

Regarding what differences in distributions the DDE statistic might be sensitive to, the general entropic nature of the statistic suggests a considerable variety of different distribution characteristics would be distinguishable. The fundamental contrast represented by DDE is between the theoretical entropies associated with the distributions encompassed by null hypotheses and the entropies associated with sample data as expressed through a nonparametric estimate of those entropies. The entropy gap represented by DDE is affected by global differences in the shape and spread of the two distributions being contrasted. For example, the statistic would be sensitive to mismatches in tail thickness and breadth of effective support, since a broad region of small density values $f(x)$ will tend to result in large aggregate contributions to entropy values $-f(x)ln(f(x))$. Another density feature that can cause larger differential entropy value differences is degree of skewness, which tends to shift the density away from its peak and alter log-density values. In particular, this would be the case when the parametric family of distributions specified by the null hypothesis is a symmetric family of distributions (e.g., Normal, t-distribution, Logistic), but the actual population density is skewed (e.g., Gamma, Weibull, and most Betas), and vice-versa. Other features of distributions that would be expected to accentuate differences in differential entropies include such things as multimodality, which can spread out density mass and increase entropy, and differing kurtosis which can change peak and tail density mass, also affecting entropy.

The entropic nature of the DDE metric as a global shape diagnostic reflects more than pointwise fit and portrays how dispersed or concentrated distributions are, as well as how density differences vary across subsets of a random variable's support. Because differential entropy is integrated over the entire relevant range of $X$, the statistic is especially sensitive to alternatives that alter global dispersion or change tail behavior. However, as for all statistical tests, not rejecting a null hypothesis because of small and/or insignificant DDE values can be the result of insufficient power against alternatives that exhibit inherently similar differential entropy profiles, e.g., a standard normal and a standard logistic distribution. Moreover, in terms of the theory underlying differential entropy, it is not necessarily the case that identical differential entropy values imply identical probability distributions – while unusual in practice for reasons noted ahead, pathological cases can occur where differently-shaped distributions can have the same differential entropy value.

The preceding cautions noted, in typical statistical applications, and in the sense of a practical heuristic, if a fitted KDE and a parametric model are similar in shape, their entropy values will also be generally similar — and vice versa. To motivate the first point, note that the differential entropy functional $DE(f) = -\int f(x)\ln(f(x))\,dx$ is continuous under mild regularity conditions on $f$, which implies that small changes in the density lead to small changes in differential entropy. Then if two densities $f$ and $\xi$ are similar in the sense that $\|f-\xi\|_1$ is small, $|DE(f)-DE(\xi)|$ will tend to be small as well. Moreover, many statistical modeling contexts involve unimodal distributions that are bounded and/or have thinning tails, and empirical density estimates are based on data with finite variances and/or bounded supports. In this setting, small differences in location, scale, skewness, or kurtosis tend to induce small shifts in *ln*-density terms, resulting in entropy values that remain close to each other. And if two distributions have similar supports, have means and variances that are also similar, and differ only modestly in terms of higher order moments, entropy values tend to be similar in value.

Reversing the direction of heuristic logic, if two models exhibit similar differential entropy values, then as a practical matter it is likely the associated density graphs will be similar, their distributional behavioral implications will also be similar, and either of the associated distributional models may be a reasonable approximation to the other for modeling purposes. In this sense, either DDE test outcome is constructive, where not rejecting $H_0$, while not necessarily suggesting the null distribution is literally true, can nevertheless suggest it mimics the true distribution. If two densities have differential entropies that are close in value and they are smooth, unimodal, and have similar supports, they are not likely to be substantially different either graphically or in the assignment of event probabilities.

The preceding discussion suggests that in typical applications, if a null distribution is not rejected because the sampling distribution has a sufficiently small DDE value, the null distribution might be considered adequate for modelling purposes even if is not literally correct. However, as a theoretical counterpoint, if in a statistical modeling context, the set of data generating processes under consideration cannot plausibly be considered to be encompassed by conventional functional families (e.g., see Appendix A), the fact remains that atypical distributions can be devised for which nearly identical differential entropies do not imply nearly identical distributions. Such cases can be further empirically investigated by direct comparisons of the densities encompassed by the null hypothesis to the KDE density representations of the sampling density.

### *5.2 Empirical Implementation of the DDE Test Statistic*

The empirical implementation of the DDE test requires that appropriate critical values or probability values associated with the test be defined which are linked to the distribution of the $DDE(f)$ statistic induced under $H_0$. The DDE statistic has an asymptotic normal distribution under general regularity conditions that apply to a wide collection of parametric families of distributions commonly implemented in statistical practice. Thus, it is conceptually possible to derive critical values or probability values with asymptotic validity based on asymptotic distributions. Moreover, a test defined in this way will be a consistent test with power converging to 1 as sample size increases. Theorems B.3 – B.5 in the Appendix provide general results relating to the asymptotic distributions of the $DDE(f)$ statistic based on familiar classical regularity conditions. The results are presented here both to support the preceding statements relating to asymptotic behaviors, and also for independent interest since while most of the components of the proofs exist in the literature, tying them together in the ways stated in the theorems do not appear readily available.

While one could identify the specific functional components that identify the asymptotic distributions of $DDE(f)$ for each distribution of interest, such as those listed in Appendix A, the implementation details are idiosyncratic to specific parametric families of distributions, and the approach requires devising efficient estimators for unknown values of parameters that define the functional forms of the asymptotic distributions. Instead of following this traditional approach to devising an asymptotically valid test, a parametric bootstrapping methodology is invoked for characterizing the distribution of the DDE test statistic and identifying critical values and probability values of the test. The reasoning for adopting this less traditional approach is provided below.

### 5.2.1 *Parametric Bootstrapping of the DDE Test Statistic*

Note the DDE statistic is fundamentally the difference between a parametric and nonparametric functional, with differing asymptotic convergence rates, despite the ultimate existence of the limiting distribution to which the statistic converges. The parametric bootstrap implemented here resamples from the estimated null distribution and recalculates DDE values, forming an empirical distribution that encompasses the small sample bias inherent in KDE estimates, the bias present in the parametric estimator, and adapts to the true underlying sampling distribution, and not just its asymptotic normal approximation. Thus, for purposes of hypothesis testing, no explicit "correction" for these biases is necessary – they are in effect already accounted for in the bootstrapped sampling distribution. In addition, entropy estimation has been shown to be relatively slow to converge to its asymptotic distribution and can be notably non-normally distributed in small to moderate-sized samples. Relatedly, for nonlinear functionals such as the DDE, bias and second-order accuracy are notably improved through use of the bootstrap approach as opposed to relying on typical asymptotic derivations and associated results (Hall and Morton 1993; and Beran 1977).

The specific steps involved in generating the parametric bootstrap samples, calculating the critical values or probability values of the DDE test, and then generating a test outcome are as follows[5]:

1) Generate $x_i^{boot} \sim iid\ f\left(x;\hat{\boldsymbol{\theta}}_{ML}\right)$, $i = 1,\cdots,n$, where $H_0 : X \sim f\left(x;\boldsymbol{\theta}\right)$
2) $\hat{\boldsymbol{\theta}}_{ML}^{boot} = \underset{\boldsymbol{\theta}\in\Omega}{\arg\max}\left(n^{-1}\ln\left(f\left(x_i^{boot};\boldsymbol{\theta}\right)\right)\right)$
3) $DE_{ML}^{boot}\left(f\right) = -\int_x f\left(x;\hat{\boldsymbol{\theta}}_{ML}^{boot}\right)\ln\left(f\left(x;\hat{\boldsymbol{\theta}}_{ML}^{boot}\right)\right)dx$ , which equals $MaxEnt\left(\hat{\boldsymbol{\theta}}_{ML}^{boot}\right)$, a plug-in value, if $f$ is a MaxEnt Distribution
4) $h_{boot} = k\left(n\right)c_{boot}\,n^{-1/5}\hat{\sigma}_{\mathbf{x}^{boot}}$ , where $c_{boot}$ and $k\left(n\right)$ are chosen as in section 4.2.3, based on the category of null distribution and bootstrap sample outcomes
5) $$\begin{Bmatrix} \hat{f}^{boot}\left(x;h_{boot}\right) \\ \hat{g}^{boot}\left(y;h_{boot}\right) \end{Bmatrix} = \begin{Bmatrix} \dfrac{1}{nh_{boot}}\sum_{i=1}^{n}K\left(\dfrac{x-x_i^{boot}}{h_{boot}}\right) \\ \dfrac{1}{nh_{boot}}\sum_{i=1}^{n}K\left(\dfrac{y-y_i^{boot}}{h_{boot}}\right) \end{Bmatrix} when \begin{Bmatrix} R\left(X\right)=\mathbb{R} \\ R\left(X\right)=\mathbb{R}_+ \wedge Y=\ln\left(X\right) \end{Bmatrix}$$

[5] In the event that the MLE is difficult or intractable to compute, the large sample convergence and bootstrap calibration of the approach follow through if a method of moments estimator is used in place of the MLE outcomes. However, analytical bias and variance expressions would need to be redefined.

6) $$DE_{KDE}^{boot}\left(\hat{f};h_{boot}\right)=\begin{cases}-\int\limits_{x\in\mathbb{R}}\hat{f}^{boot}\left(x;h_{boot}\right)\ln\left(\hat{f}^{boot}\left(x;h_{boot}\right)\right)dx\\-\int\limits_{y\in\mathbb{R}}\hat{g}^{boot}\left(y;h_{boot}\right)\ln\left(\hat{g}^{boot}\left(y;h_{boot}\right)\right)dy+\overline{y}_{boot}\end{cases}\text{ when }\begin{cases}R\left(X\right)=\mathbb{R}\\R\left(X\right)=\mathbb{R}_{+}\wedge Y=\ln\left(X\right)\end{cases}$$

7) Repeat 1)-6) $n_{boot}$ times to generate a bootstrapped distribution of differential entropy values $\mathbf{DDE}^{boot}\left(f;h_{boot}\right)=\left\{DDE^{boot}\left(f;h_{boot}\right)_{j}\ ,j=1,\cdots,nboot\right\}$.

8) Define $\alpha-level$ critical values or *p*-values for the test based on the bootstrapped distribution as $Q_{\alpha/2}^{boot}\left(\mathbf{DDE}^{boot}\left(f;h_{boot}\right)\right)$ and $Q_{(1-\alpha/2)}^{boot}\left(\mathbf{DDE}^{boot}\left(f;h_{boot}\right)\right)$, where $Q_{\tau}^{boot}$ refers to quantiles, and

$$p-value=\frac{1+\sum_{b=1}^{nboot}1\left(\left|DDE^{boot}\left(f;h_{boot}\right)-\overline{DDE}\right|\geq\left|DDE\left(f\right)-\overline{DDE}\right|\right)}{nboot+1}.$$

9) Reject $H_0$ at the $\alpha-level$ of Type I Error if $DDE\left(f\right)\notin\left[Q_{\alpha/2}^{boot},Q_{(1-\alpha/2)}^{boot}\right]$ or $p-value\leq\alpha$.

Note, the +1 in the equation for calculating the *p*-value is the standard *plus-one* Monte Carlo permutation adjustment for the test to behave correctly for finite *nboot*, i.e., it is consistent with an implementation of the exact finite-simulation rank test.

Arguably one of the most important benefits of implementing the parametric bootstrap methodology is bias mitigation. It is known that differential entropy estimators—including those based on the MLE and KDE— exhibit finite-sample bias as noted in sections 4.1.1 and 4.2.2. The MLE of entropy tends to be negatively biased while KDE-based estimators typically exhibit positive or negative bias depending on the choice of bandwidth, kernel, and underlying distributional curvature (Joe 1989; Delyon and Portier 2016). These biases can be especially notable for small to moderate sample sizes, and they can complicate the use of entropy as a test statistic for distributional inference.

To account for and mitigate the impact of these biases, the parametric bootstrap approach calibrates the null distribution of the entropy-based test statistic, inclusive of biases. The key insight regarding bias mitigation is the bias-canceling property of the parametric bootstrap. The MLE and KDE of differential entropy have first order stochastic expansions under typical regularity conditions (see Theorems B.3-B.5) as

$$\left\{\begin{array}{c} DE_{ML}\left(f\right) \\ DE_{KDE}\left(\hat{f};h\right)\end{array}\right\} = DE\left(f\right) + \left\{\begin{array}{c} Bias_{MLE}\left(n\right) \\ Bias_{KDE}\left(\hat{f};h\right)\end{array}\right\} + o_p\left(1\right). \tag{19}$$

Then under $H_0$, $DDE\left(f\right) = Bias_{MLE}\left(n\right) - Bias_{KDE}\left(\hat{f};h\right) + o_p\left(1\right)$, which reveals that $DDE\left(f\right)$ is not generally centered at zero unless the bias terms were to cancel exactly, underscoring that relying on asymptotic normality alone to approximate the distribution of $DDE\left(f\right)$ would lead to miscalibrated tests and distorted inferences.

The parametric bootstrap mimics the sampling distribution of the test statistic inclusive of bias, where $DDE^{boot}\left(f\right) = Bias_{MLE}^{boot}\left(n\right) - Bias_{KDE}^{boot}\left(\hat{f};h\right) + o_p\left(1\right)$. Provided that $\hat{f}$ consistently estimates $f$, the bootstrapped biases converge to the sample data-based biases, with the remaining $o_p\left(1\right)$ term reflecting the stochastic variability of the sample data. Then under $H_0$

$DDE^{boot}\left(f\right) \overset{d}{\approx} DDE\left(f\right)$, the approximation holding in probability, and the bootstrap distribution reflects both the bias and variance structure of the DDE statistic. Using the quantiles of $DDE^{boot}\left(f\right)$ to determine critical values or calculating $p-values$ for hypothesis testing ensures that the effect of bias is accounted for by appearing symmetrically in both the observed and simulated statistics. In effect, the parametric bootstrap inherently adjusts for finite sample biases of the two estimators of differential entropy. No explicit bias corrections are needed, and the method inherently adapts to data-driven bandwidths and KDEs applied to transformed data (e.g., log-scale KDE based estimation for random variables with nonnegative support). This ensures that test calibration and Type I error control remain valid even if bandwidths are not perfectly tuned (Efron and Tibshirani 1993). The overall validity of the approach relies on well-established asymptotic results relating to bootstrapping methodology (Berrett et al. 2019).

## 6. Simulations of Finite Sample Size and Power of the DDE Test

The DDE test is simulated in this section to illustrate and evaluate its empirical size and power across a range of null and alternative distributions. The four null hypotheses examined correspond to parametric distribution families commonly used in practice and align with the categories discussed in section 4.3.2: the Normal, Exponential, Gamma, and LaPlace distribution families. The standard Normal, $N\left(0,1\right)$, is used as the data-generating process for assessing the DDE test under symmetric distributions with support on the entire real line. The $Exponential(2)$ distribution (equivalently, $Chisquare(2)$) is implemented to study test performance for highly right-skewed distribution families with nonnegative and exponentially declining support. The $Gamma(3,1)$ distribution is used to analyze test behavior under right-skewed, unimodal distributions with nonnegative support and long slowly diminishing right tails. Finally, the $LaPlace\left(0,1/\sqrt{2}\right)$ distribution represents a cusped distribution with excess kurtosis substantially higher than that of the Gaussian. The alternative distributions used to study the size and power of the DDE test are listed in Table 4.

**Table 4. Simulated Null Hypotheses and Alternative Distributions**

| $H_0$: **Family** | $H_0$: **Member** | **Alternative Distributions** | | | |
|---|---|---|---|---|---|
| $N(u,\sigma^2)$ | $N(0,1)$ | $LaPlace\left(0,\sqrt{.5}\right)$ | $Logistic\left(0,\frac{\sqrt{3}}{\pi}\right)$ | $Cauchy(0,1)$ | $\frac{t(3)}{\sqrt{3}}$ |
| $Exp(\theta)$ | $Exp(2)$ | $Rayleigh\left(\left(\frac{8}{4-\pi}\right)^{1/2}\right)$ | $LogLogistic(2.6954,\ 1.5764)$ | $Lomax(3,4)$ | $LogNormal(.3466, ln(2))$ |
| $Gamma(\alpha,\beta)$ | $Gamma(3,1)$ | $Weibull(1.7915,\ 3.3727)$ | $LogLogistic(3.72,\ 2.66)$ | $InvGaussian(3,9)$ | $LogNormal\left(.9548, ln\left(\frac{4}{3}\right)\right)$ |
| $LaPlace(u,b)$ | $LaPlace\left(0,\sqrt{.5}\right)$ | $N(0,1)$ | $Logistic\left(0,\frac{\sqrt{3}}{\pi}\right)$ | $Cauchy(0,1)$ | $\frac{t(3)}{\sqrt{3}}$ |

The parameters of the non-null (alternative) simulated distributions were chosen to match the variances of the corresponding null distributions, which were 1, 4, 3 and 1 for the Normal, Exponential, Gamma, and LaPlace distributions, respectively. The exception was the Cauchy distribution, for which moments do not exist. Where feasible, the respective means of the null distributions – 0, 2, 3, and 0 – were also matched. Exceptions were the Rayleigh and LogLogistic distributions used as alternatives for the Exponential null, which have means of 3.8261 and 3.4641, respectively. The chosen distributions span a range of shapes, spreads, skewness levels, and degrees of similarity to the null distributions, and are shown graphically in Figure 1.

**Figure 1: Comparison of Null and Alternative Distributions**

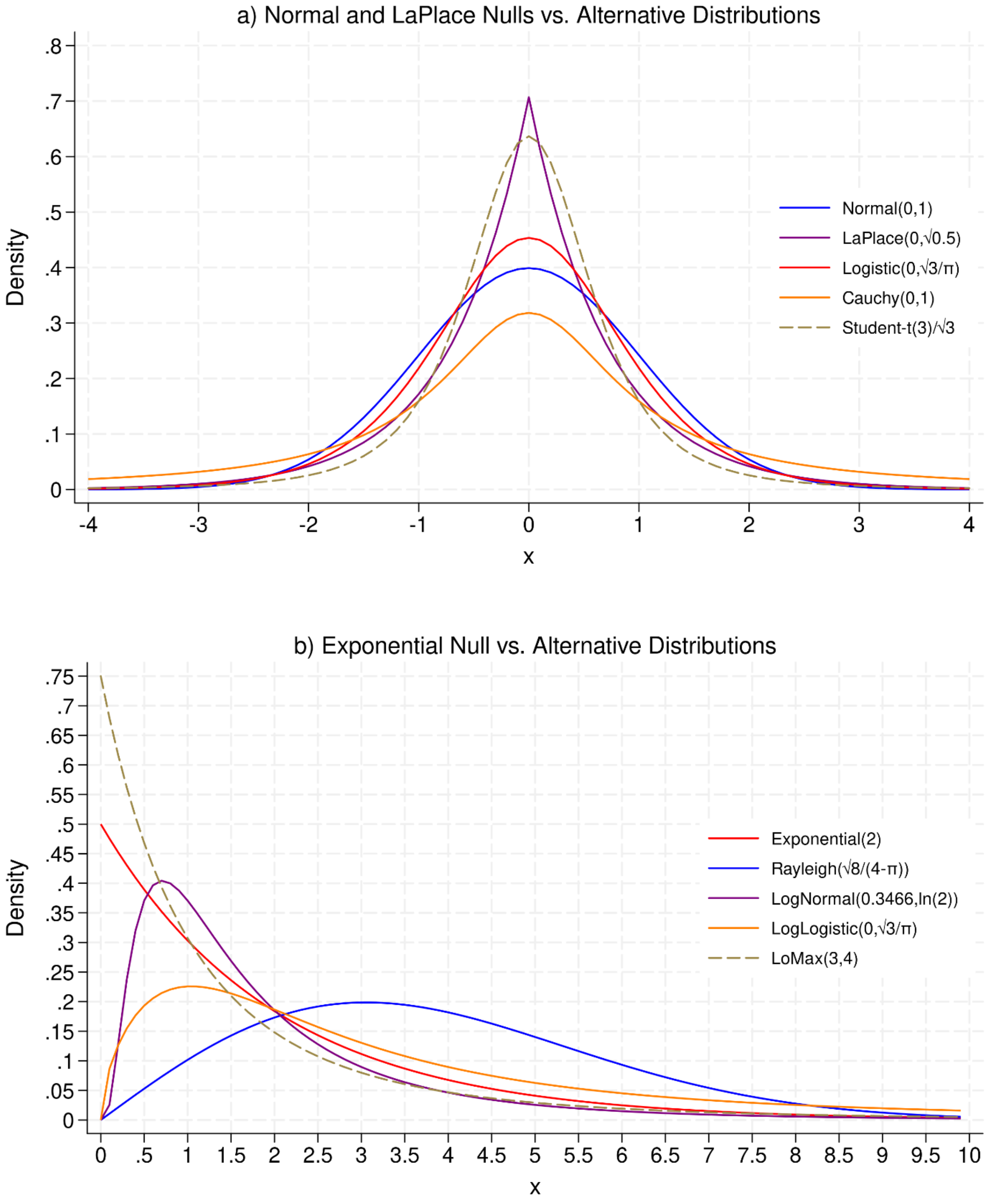

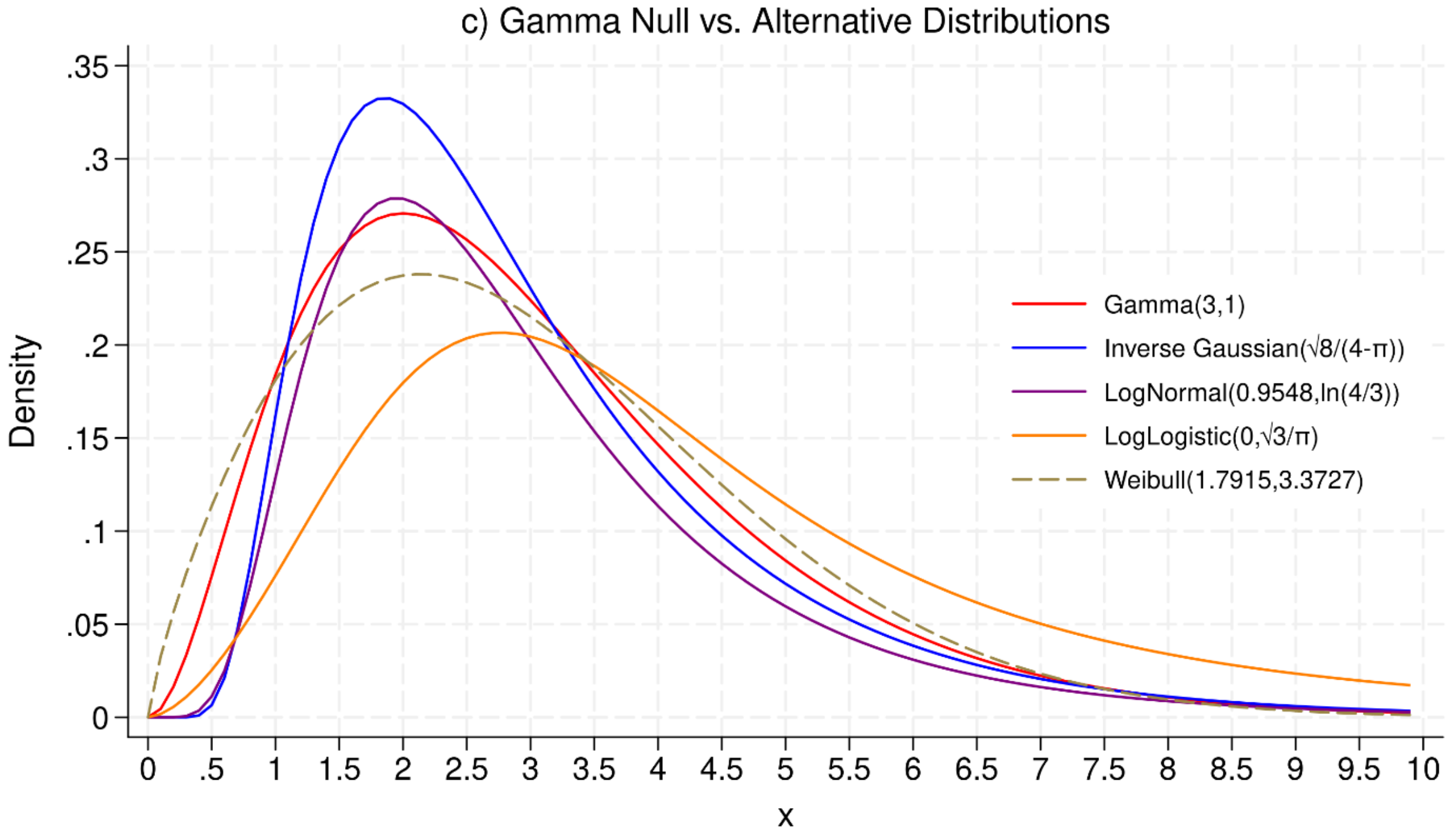

Simulated sample sizes of n = 50, 100, 250 and 500 were used to assess the size and power of the DDE test, spanning small to moderately large data sets. The bootstrapped distributions were generated using $n_{boot} = 1000$ bootstrap repetitions, and the number of Monte Carlo repetitions used to evaluate test size and power was likewise set to 1000. The probability of Type I error for testing the null hypotheses was set at the often used $\alpha = .05$ level.

Figure 2 displays the empirical sizes of the DDE test for the four simulated null distributions. The results indicate that the DDE testing procedure maintains reasonably tight control of Type I error probabilities even at relatively small sample sizes, highlighting one of the advantages of implementing the test through parametric bootstrapping. The graphs illustrate the consistency of the DDE test, with power increasing as the sample size $n$ grows for all cases considered. As expected, the rate at which power increases with sample size depends on the characteristics of the alternative distribution. Consistent with the discussion in section 5.1, the power of the DDE testing procedure is substantial against alternative distributions that are notably different from the null distribution, and in these cases, power is appreciable even for small sample sizes. Also, in line with section 5.1, the DDE test exhibits less power against sampling distributions that closely resemble the null in shape and probability assignments. In particular, the alternative distributions associated with the lowest power include the Logistic distribution under the Normal null, the LoMax distribution under the Exponential null, the Weibull distribution under the Gamma null, and the Student-t distribution under the LaPlace null. Graphs highlighting these specific null-alternative pairs are presented in Figure 4.

**Figure 2: Empirical Sizes of DDE test**

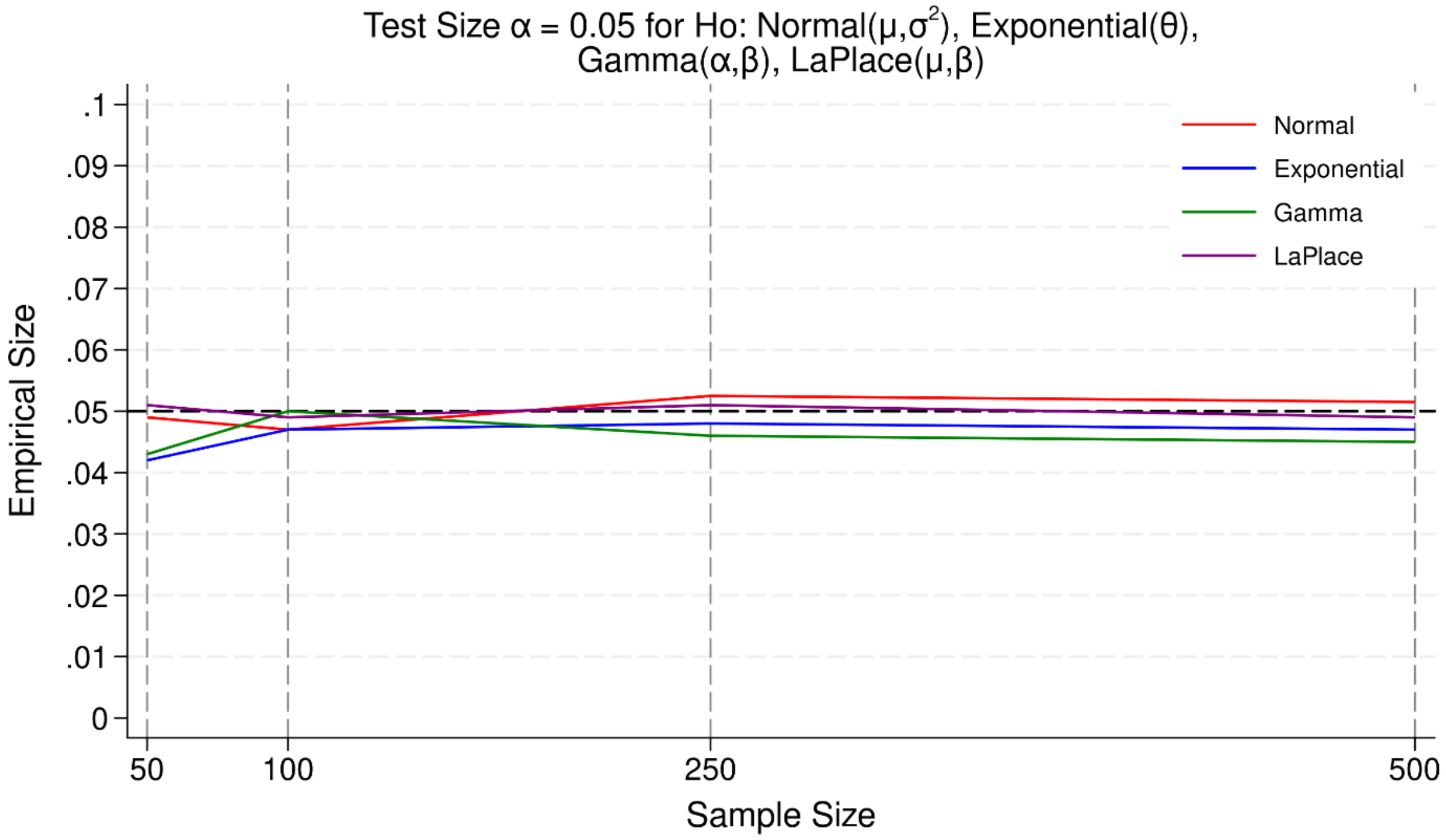

The power of the DDE test against the alternative distributions presented in Table 4 and depicted in Figure 1 is shown in Figure 3. While certainly not perfect substitutes, the graphs in Figure 4 reinforce the observations made in section 5.1 regarding the constructive nature of the DDE test for either test outcome. In particular, cases in which the null distribution is not rejected – even though this is literally a Type II error – can nevertheless indicate that the null distribution closely approximates the true data-generating process. In such situations, the null and true distributions do not differ substantially in shape and probabilistic behavior, and the magnitude of any errors arising from analyses that proceed under the null distribution are attenuated in these cases.

**Figure 3: Empirical Power of DDE Tests**

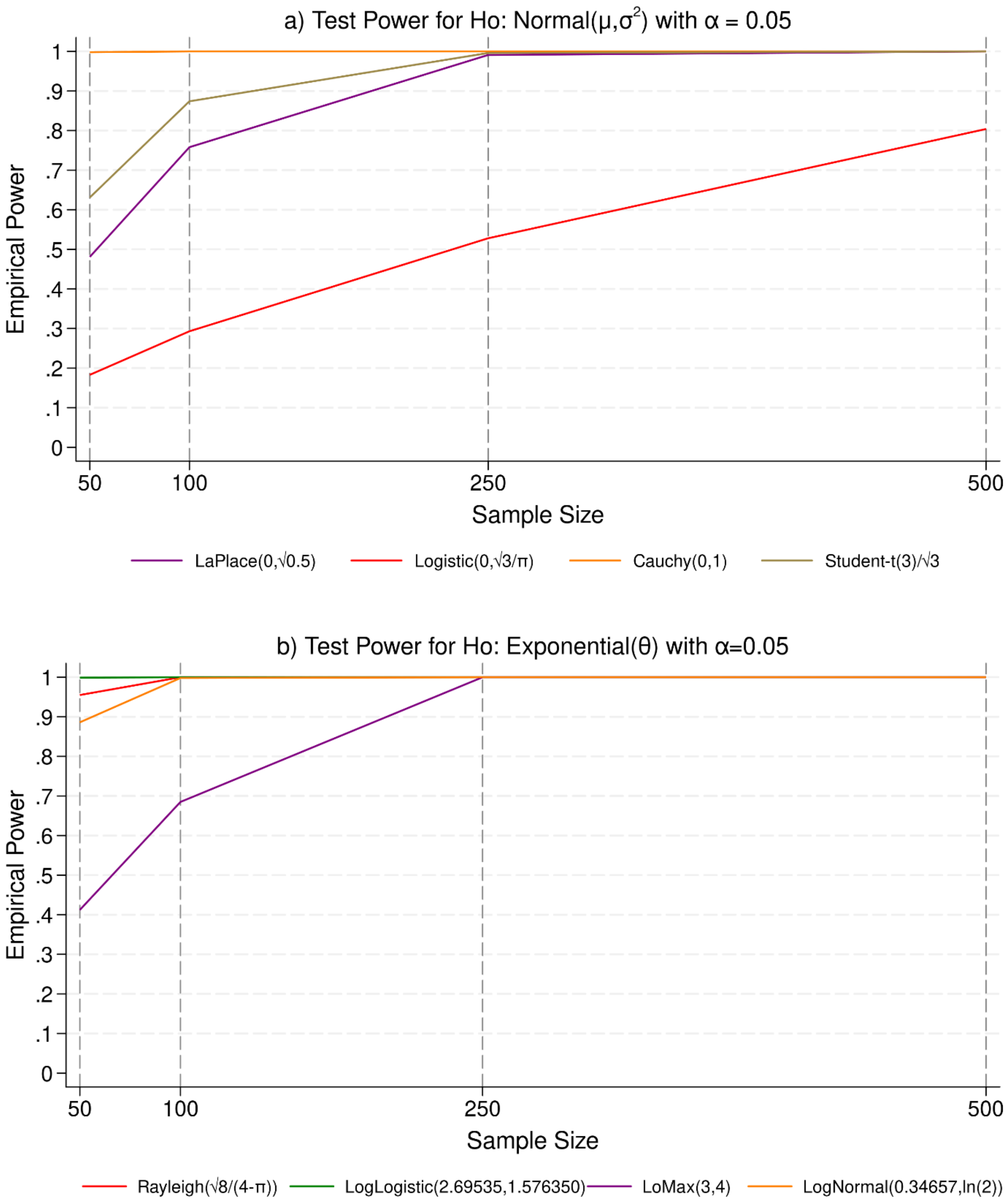

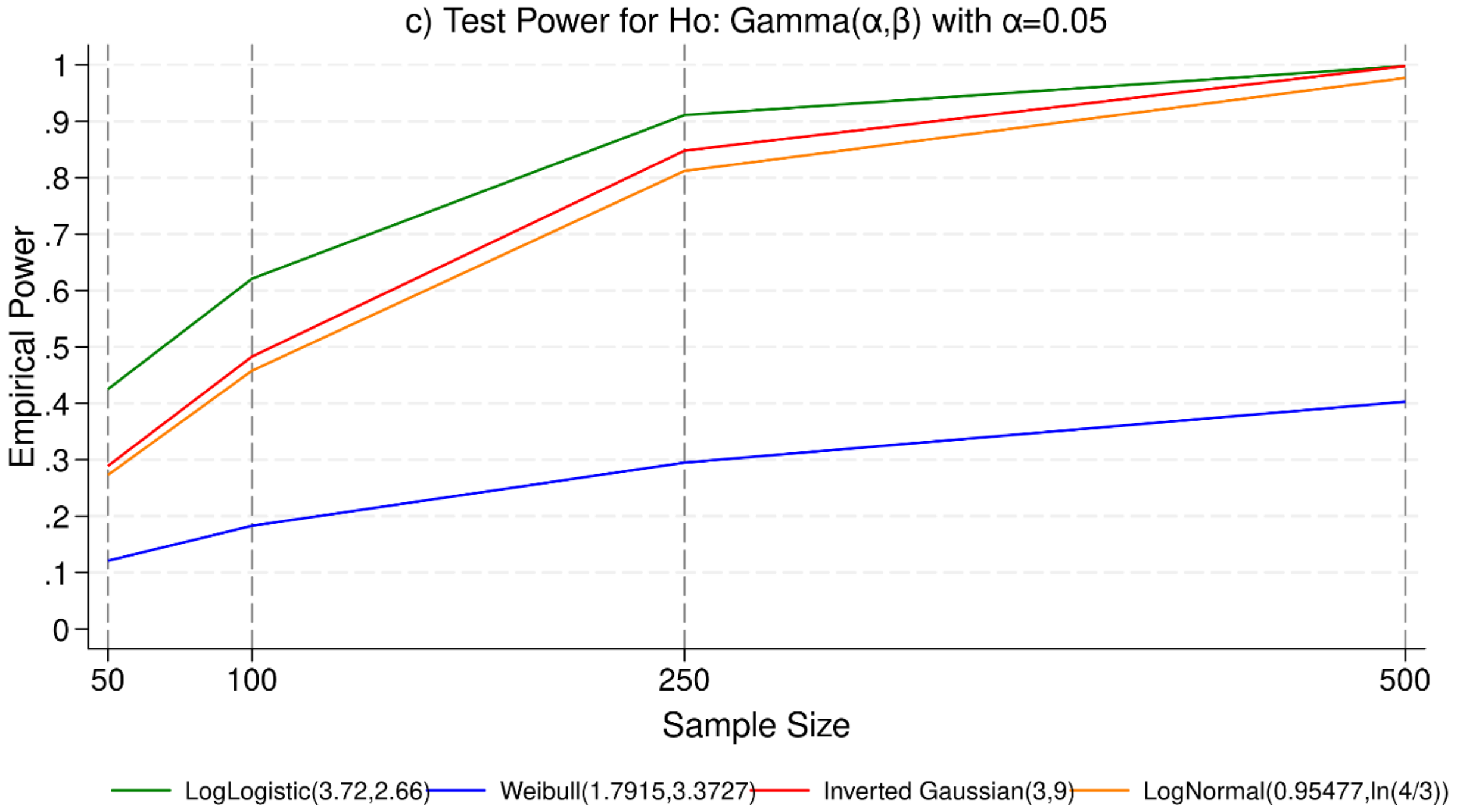
c) Test Power for Ho: Gamma(α,β) with α=0.05
Empirical Power
1
.9
.8
.7
.6
.5
.4
.3
.2
.1
0
50
100
250
500
Sample Size
LogLogistic(3.72,2.66)
Weibull(1.7915,3.3727)
Inverted Gaussian(3,9)
LogNormal(0.95477,ln(4/3))

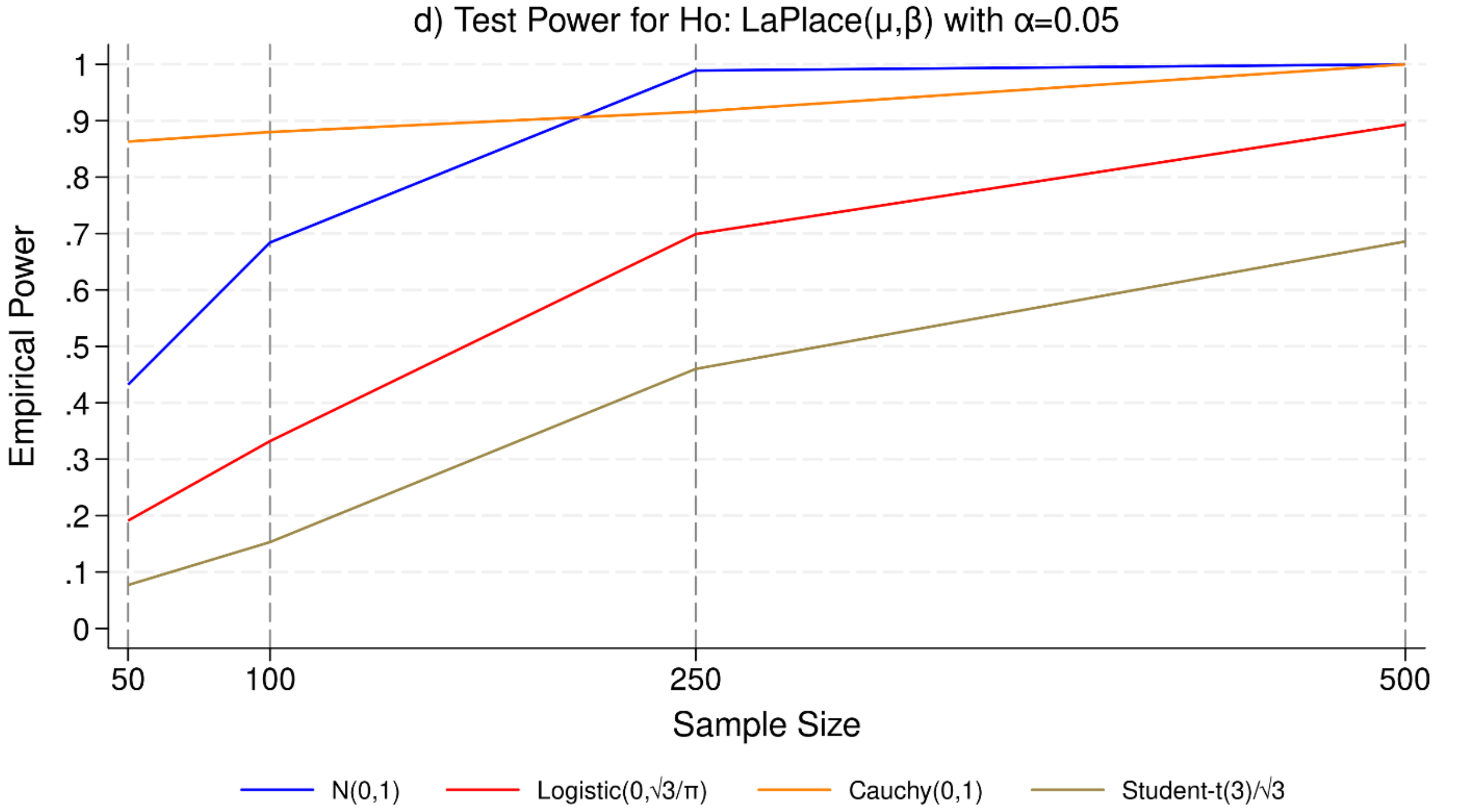
d) Test Power for Ho: LaPlace(μ,β) with α=0.05
Empirical Power
1
.9
.8
.7
.6
.5
.4
.3
.2
.1
0
50
100
250
500
Sample Size
N(0,1)
Logistic(0,√3/π)
Cauchy(0,1)
Student-t(3)/√3

**Figure 4. Comparison of Null with Lowest Power Alternative Distributions**

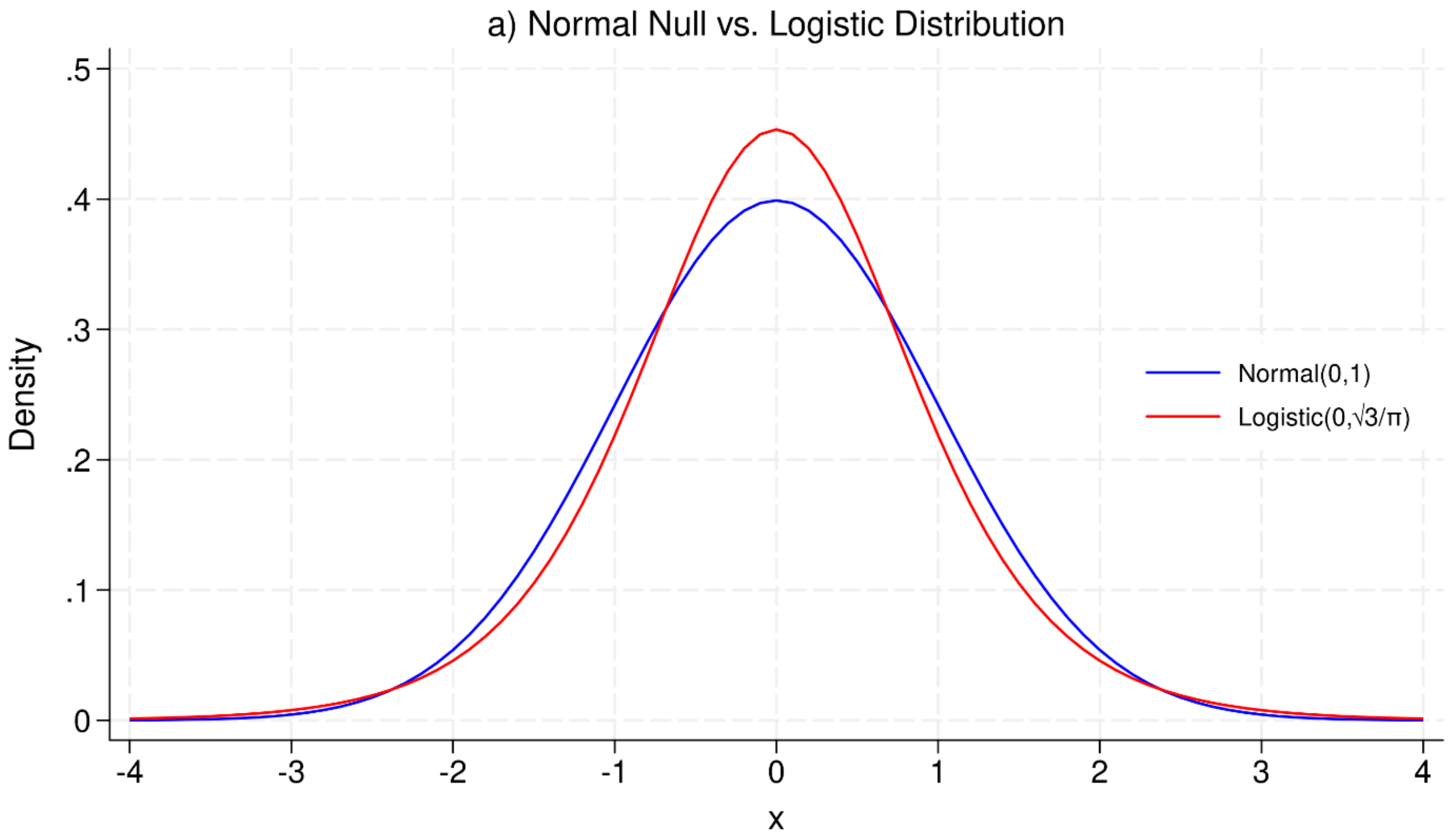

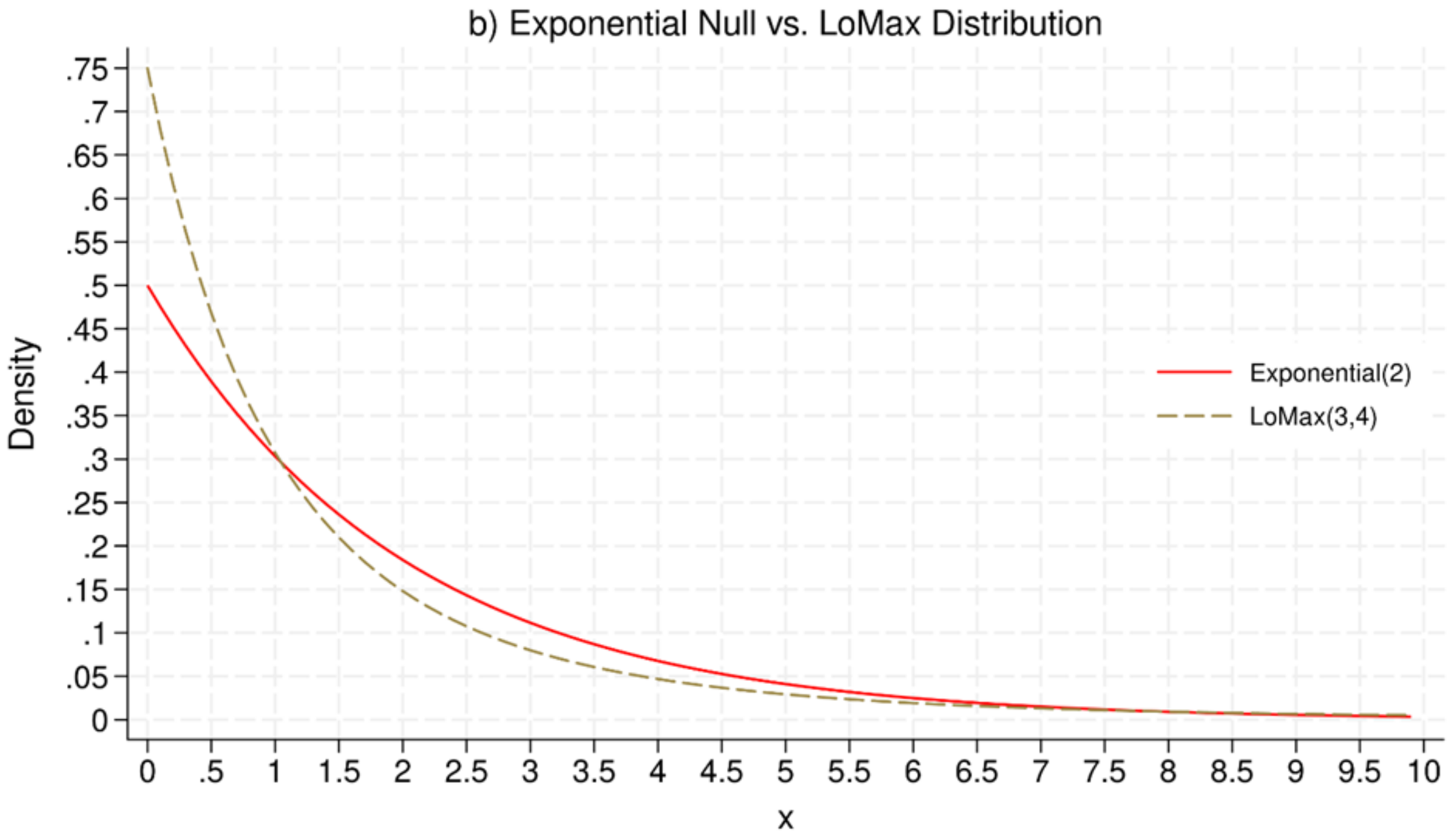

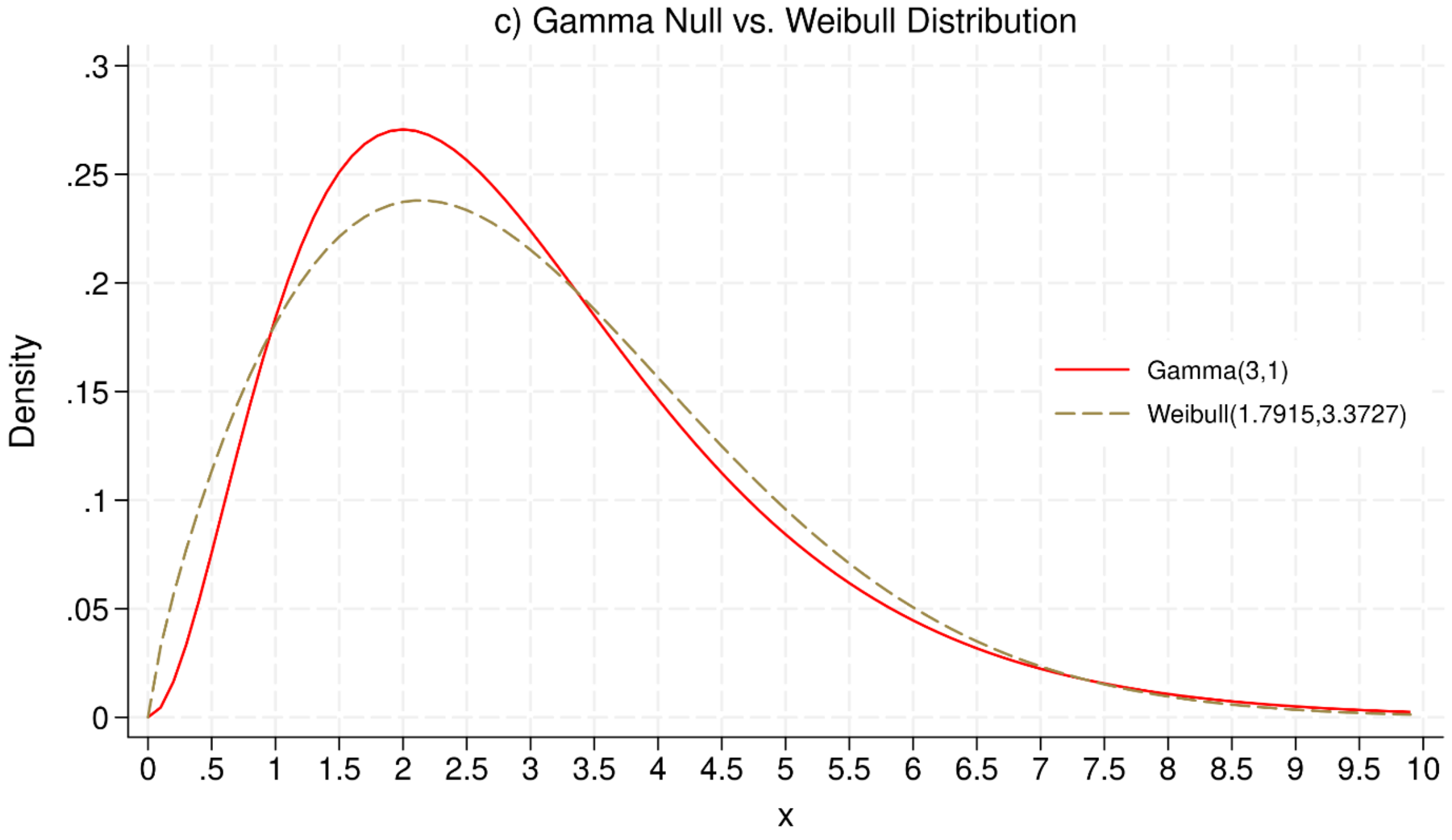

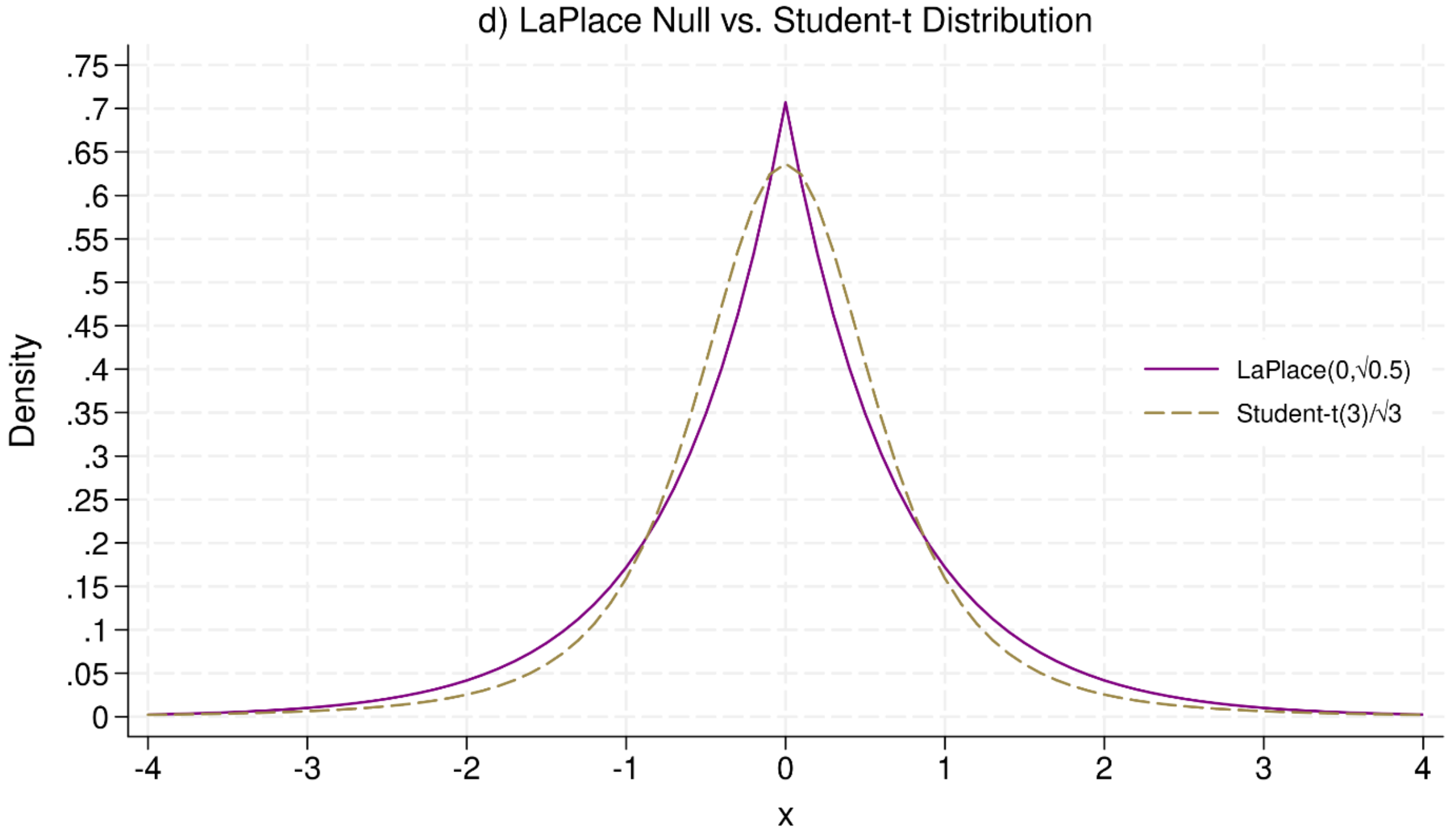

## 7. Empirical Applications

In this section, we apply the DDE testing methodology to three classical data sets that have been widely examined in the literature and span distinct empirical domains. The first consists of two well-known environmental duration data sets relating to the waiting times between eruptions of the Old Faithful Geyser, as analyzed by Härdle (1991) and Azzalini and Bowman (1990). The second pertains to actuarial and risk-management data on the incidence and loss severities of Danish Fire Insurance claims, originally presented by Embrechts et al. (1997). The third is an application of econometrics and industrial organization involving the residuals from Christensen and Greene's (1976) Translog cost function analysis of the U.S. electrical power industry. All DDE tests reported here are based on bootstrap sample sizes of $n_{boot} = 1000$, with all other methodological details as described previously.

### 7.1 Old Faithful Geyser Waiting Times Between Eruptions

This application concerns observations on the waiting times between eruptions of the Old Faithful Geyser in Yellowstone National Park. The Azzalini and Bowman (1990) dataset contains 299 paired measurements on the Old Faithful geyser, indicating eruption waiting times to the next eruption, measured in minutes continuously from August 1 to August 15, 1985. In contrast, the dataset presented by Härdle (1991) contains 272 paired observations and is widely used in classroom and tutorial settings for illustrating waiting time events. According to the documentation in the R programming language, which uses both data sets for empirical illustrations "[t]here are many versions of this dataset around: Azzalini and Bowman (1990) use a more complete version."[6] Given the waiting times nature of the data, probability distribution families that have often been found to be useful in such contexts include the Gamma and Lognormal parametric families. If either provided a satisfactory fit, they would offer a straightforward and familiar probability model characterization of the eruption waiting times. A third parametric family considered as a null distribution family is the three-parameter Generalized Gamma family, expressed in terms of Stacy's (1962) parameterization (see Appendix A). This family is a very flexible collection of distributions that subsumes the Gamma, Exponential, Weibull, Rayleigh, Nakagami, and Half-Normal as special subfamilies. Finally, although not expected to be an appropriate probability distribution for these data, the ubiquitous Normal family of distributions was also evaluated as a null distribution in a possibly approximative sense. The DDE test results for each of these null hypotheses were as follows:

[6] https://search.r-project.org/R/refmans/datasets/html/faithful.html

$$H_0:\left\{\begin{matrix}\left\{\begin{matrix}Gamma(\alpha,\beta)\\ Lognormal(u,\sigma^2)\\ Normal(u,\sigma^2)\\ Generalized\ Gamma(a,d,p)\end{matrix}\right\}\\ \left\{\begin{matrix}Gamma(\alpha,\beta)\\ Lognormal(u,\sigma^2)\\ Normal(u,\sigma^2)\\ Generalized\ Gamma(a,d,p)\end{matrix}\right\}\end{matrix}\right\} \Rightarrow p-value=\left\{\begin{matrix}\left\{\begin{matrix}0\\0\\0\\.947\end{matrix}\right\}\\ \left\{\begin{matrix}0\\0\\0\\.733\end{matrix}\right\}\end{matrix}\right\}\ for\ \left\{\begin{matrix}\{H\ddot{a}rdle\}\\ \{Azzalini\ and\ Bowman\}\end{matrix}\right\}\ Data$$

Thus, the familiar Gamma, Lognormal, and Normal parametric probability models are all decisively rejected by the DDE tests when applied to Old Faithful Geyser eruption data, based on either of the two available data sets. On the other hand, the substantially more flexible Generalized Gamma parametric family is not rejected. These results suggest that less flexible distribution families such as Gamma and Lognormal are insufficient to model eruption waiting times, and a more complex probability model is likely warranted to explain the probability of their behavior.

### 7.2 Danish Fire Insurance Loss Severities

This dataset was collected by the Copenhagen Reinsurance Company Ltd. and consists of 2,167 fire insurance loss claims recorded between 1980 and 1990. The loss amounts were adjusted for inflation to 1985 price levels and are expressed in millions of Danish Krone. The data set has become a standard benchmark in teaching and research applications within Extreme Value Theory and has appeared in numerous textbooks on risk and extreme value analysis in finance, including Embrechts et al. (1997) and McNeil et al. (2015). Given the heavily right-skewed nature of the data, the highly flexible Generalized Gamma family was used as one of the null distributions to be tested. As noted in the previous section, this family can represent a wide range of right skewed/heavy-tailed distributional forms and is thus a reasonable candidate for a single-distribution fit to these data. Simpler parametric distributions commonly used for modeling right-skewed data, namely the Gamma and the Lognormal distributions, were also tested. As in section 7.1, although not expected to provide an appropriate probability distribution for this data, the ubiquitous Normal family of distributions was again included as a null distribution hypothesis for comparison in a possibly approximative sense. The DDE test results for each of these null hypotheses are summarized as follows:

$$H_0 : \left\{ \begin{array}{c} Gamma\left(\alpha,\beta\right) \\ Lognormal\left(u,\sigma^2\right) \\ Normal\left(\left(u,\sigma^2\right)\right) \\ Generalized\ Gamma\left(a,d,p\right) \end{array} \right\} \Rightarrow p-value = \left\{ \begin{matrix} 0 \\ 0 \\ 0 \\ 0 \end{matrix} \right\}$$

In this case, all of the parametric families of distributions tested were decisively rejected by the DDE test, including the substantially more flexible Generalized Gamma parametric family. Given in particular the rejection of the Generalized Gamma family, the implication is that the host of simpler nested probability models noted in section 7.1 are also decisively rejected. In this application, the results suggest that the search for an appropriate probability model for these insurance loss severities requires exploration of additional, more flexible and likely more complex probability models.

### 7.3 Translog Cost Function

This data set consists of two sets of estimated residuals derived from the Translog cost functions estimated by Christensen and Greene (1976) in their analysis of economies of scale in U.S. electric power generation. One set of residuals corresponds to the unrestricted model, while the other set incorporates homogeneity restrictions. Each set includes three estimated residual vectors: one for the Translog cost function, $\ln(C)$, and two for the associated share equations for capital $(S_K)$ and labor $(S_L)$.[7] The equations were estimated jointly as a system using iterated Seemingly Unrelated Regression (SUR), which converges to MLE. The maintained assumption in their work is that residuals are jointly Normally distributed—an assumption consistent with the estimation procedure implemented (iterated SURE or ML). Under standard Central Limit Theorem (CLT) regularity conditions, residuals from such estimated models tend toward Normality as sample size increases. The DDE test results for each set of residuals, under the null hypothesis that the residuals follow a Normal parametric family of distributions, are summarized in Table 5.

[7] The fuel share equation was omitted from estimation because the cost shares necessarily sum to one, creating a singular covariance matrix; hence, one share equation must be dropped from the system.

**Table 5. DDE Tests of Translog Cost Function Residuals for $H_0 : N\left(u, \sigma^2\right)$**

| | Unconstrained Equation | | | Constrained Equation | | |
|---|---|---|---|---|---|---|
| Equation | $\ln(C)$ | $S_K$ | $S_L$ | $\ln(C)$ | $S_K$ | $S_L$ |
| DDE Test *p*-value | 0.075 | 0.000 | 0.051 | 0.229 | 0.000 | 0.050 |

The results suggest that the null hypothesis of Normality of the residuals would not be rejected at the conventional 0.05 level of Type I error for the $\ln(C)$ equation, and is at the cusp of non-rejection for the $S_L$ equations in both the unconstrained and constrained versions of the Translog cost function model. In contrast, the hypothesis of Normality is decisively rejected for both the unconstrained and constrained estimates of the $S_K$ equation. However, even for the two equations where Normality was not rejected, the associated *p*-values are small enough in three of the four cases to cast doubt on the strength of the Normality assumption. Overall, these results raise questions regarding the underlying Normality hypothesis itself, the sample size ($n = 123$) which may not have been sufficiently large to effectively induce a reliable CLT-based Normal distribution approximation, or the possibility of model misspecification.

## 8. Summary and Conclusions

This study presented a broadly applicable entropy–difference test for assessing the validity of hypotheses about which parametric distribution family generated a random sample of data. In addition to testing entire parametric families, the method is straightforwardly adaptable to assessing simple hypotheses, e.g., $H_0 : X \sim N(0,1)$. The test procedure provides a single unified testing principle that applies to a wide range of distribution families, with asymptotic validity grounded in standard maximum likelihood, bootstrapping, and kernel density estimation (KDE) principles. The method is readily implemented and straightforward to interpret. It also incorporates automated choices of bandwidth for the KDE, designed to adapt to the support and shape characteristics of the hypothesized null distribution family being tested.

The test is constructive in the sense that, whether or not the null hypothesis is rejected, the test outcome is informative: rejection implies that the proposed parametric family is inconsistent with the data, while non-rejection suggests that the hypothesized distribution is likely "close" to the actual distribution underlying the data generating process in terms of shape and probability characteristics.

The central idea is to compare the parametric ML estimate of differential entropy – reflecting the assumed null model – with a nonparametric KDE-based entropy estimate that captures the empirical distributional shape. Under the null, these two entropy estimators agree up to sampling variability, while under alternatives, they differ systematically. This results in the entropy difference being a broadly applicable goodness-of-fit measure grounded in information theory rather than moment-or likelihood-based discrepancies.

Methodologically, the study addressed several challenges inherent to entropy estimation, particularly for skewed or bounded distributions. First, to stabilize the KDE on positive supports, density estimation was performed on the *ln*-transformed scale $Y = ln(X)$, with the final entropy recovered via the appropriate transformation identity $h_X = h_Y + E[Y]$. Second, an automated and adaptive bandwidth selector was introduced to align the degree of smoothing with the shape and smoothness of the null distribution. For non-Gaussian distributions on $\mathbb{R}$ and for positive-valued right-skewed unimodal null distributions on $\mathbb{R}_+$, shape and smoothness were quantified using a kurtosis–based index, where the kernel bandwidth $h \propto \hat{\sigma}\, n^{-1/5}$ was scaled via $c = 1 + 0.1\log_2\left(\hat{\gamma}_{\text{KURT}}\right)$, where $\hat{\gamma}_{\text{KURT}}$ is a measure of the kurtosis of the null distribution relative to the kurtosis inherent in the observed data. For Gaussian and near-Gaussian distributions, no scaling was implemented, so that $c = 1$. Numerical integration was implemented via adaptive quadrature, with integration ranges for the entropy functional determined using quantile-based intervals on the appropriate numerical scale (log scale on $\mathbb{R}_+$, raw scale on $\mathbb{R}$), thereby enhancing numerical stability and preventing tail over-smoothing.

Monte Carlo studies were carried out for four representative null families—Normal, Exponential, Gamma, and LaPlace—each tested against four alternative distributions chosen to stress the test in terms of skewness, modality, tail behavior, and smoothness deviations. Small-sample behavior in the case of the non-Gaussian and right-skewed distributions (particularly at $n$ = 50) benefited from modest bandwidth inflation, whereas sample sizes of $n \geq 100$ required no such adjustment. Across all null distributions and sample sizes, empirical test sizes were relatively close to the target 0.05 level, confirming that the entropy gap statistic yields a reliably-sized test when paired with the proposed adaptive methodology. Power analyses demonstrated that the entropy–gap test was reactive to a broad range of alternatives, particularly those involving changes in smoothness, shifts in tail weight, or skewness mismatches. Notably, the test exhibited significant power for detecting deviations for Gamma and Laplace nulls—settings where many classical goodness-of-fit tests perform poorly. These findings support the entropy-gap test as being a distributionally flexible and information-theoretically grounded measure of discrepancy between data and a proposed parametric null model.

Several avenues for future research emerge from this work. Extending the method to multivariate settings would leverage existing multivariate entropy estimators but would require additional assessment of bandwidth selection and attention to curse-of-dimensionality issues. Further theoretical analysis of the asymptotic distribution of the entropy gap under particular null distributions could motivate refinements to bootstrap calibration that improve computational efficiency and accuracy. Investigating performance of the DDE approach under multimodal heavy-tailed null distributions may also suggest adjustments to the KDE component of the procedure that might enhance size and power performance. Relatedly, adopting local or variable-bandwidth KDEs may lead to improvements in sensitivity to subtle local departures from the null. Moreover, although the automated bandwidth methodology generally aligned with recommendations in the literature and performed well in simulations, further research may yield refinements that further reduce the need for case-specific tuning. Finally, given that bootstrapped DDE test statistics are used in the definition of critical and probability values for the DDE test and they are subject to sampling error, additional investigation of the effects of bootstrap sample size may result in refinements relating to the accuracy of those definitions.

Overall, this research contributes a general, flexible, and empirically validated framework for entropy-based goodness-of-fit testing. Grounded in robust nonparametric estimation methods and principled calibrations, the entropy–gap test adds to the statistical analysis toolkit, particularly for evaluating distributional assumptions where smoothness, tail structure, or skewness are central to model validity and inference.

## APPENDICES

### Appendix A. Maximum Entropy Distributions Implied by Moment Conditions and Supports

| PDF | $f(x;\Theta)$ | Support | Moment Conditions | Differential Entropy Value |
|---|---|---|---|---|
| Uniform | $(b-a)^{-1}$ | $[a,b]$ | None | $\ln(b-a)$ |
| Beta | $\frac{x^{\alpha-1}(1-x)^{\beta-1}}{B(\alpha,\beta)}$ | $(0,1)$ | $E(\ln(X))=[\psi(\alpha)-\psi(\alpha+\beta)]$ <br> $E(\ln(1-X))=[\psi(\beta)-\psi(\alpha+\beta)]$ | $\ln(B(\alpha,\beta))-(\alpha-1)[\psi(\alpha)-\psi(\alpha+\beta)]$ <br> $-(\beta-1)[\psi(\beta)-\psi(\alpha+\beta)]$ |
| Cauchy | $\frac{1}{\pi(1+x^2)}$ | $(-\infty,\infty)$ | $E(\ln(1+x^2))=2\ln(2)$ | $\ln(4\pi)$ |
| Chi-square | $\frac{1}{2^{v/2}\Gamma(v/2)}x^{(v/2)-1}\exp\left(-\frac{x}{2}\right)$ | $(0,\infty)$ | $E(X)=v$ <br> $E(\ln(X))=\psi(v/2)+\ln(2)$ | $\ln(2\Gamma(v/2))+(1-(v/2))\psi(v/2)+\frac{v}{2}$ |
| Erlang | $\frac{\lambda^k}{(k-1)!}x^{k-1}\exp(-\lambda x)$ | $(0,\infty)$ | $E(X)=\frac{k}{\lambda}$ <br> $E(\ln(X))=\psi(k)-\ln(\lambda)$ | $(1-k)\psi(k)+\ln\left(\frac{\Gamma(k)}{\lambda}\right)+k$ |
| Exponential | $\frac{1}{\theta}\exp\left(-\frac{x}{\theta}\right)$ | $(0,\infty)$ | $E(X)=\theta$ | $1+\ln(\theta)$ |
| Gamma | $\frac{1}{\beta^\alpha\Gamma(\alpha)}x^{\alpha-1}\exp\left(-\frac{x}{\beta}\right)$ | $(0,\infty)$ | $E(X)=\alpha\beta$ <br> $E(\ln(X))=\psi(\alpha)+\ln(\beta)$ | $\ln(\beta\Gamma(\alpha))+(1-\alpha)\psi(\alpha)+\alpha$ |
| Generalized Beta, 2nd Kind | $\frac{a}{bB(p,q)}\frac{(x/b)^{ap-1}}{(1+(x/b)^a)^{p+q}}$ | $(0,\infty)$ | $E\left(\ln\left(\frac{X}{b}\right)\right)=\frac{1}{a}(\psi(p)-\psi(q))$ <br> $E\left(\ln\left(1+\left(\frac{X}{b}\right)^a\right)\right)=\psi(p+q)-\psi(q)$ | $\ln(b/a)+\ln(B(p,q))+(p+q)\psi(p+q)$ <br> $-(p-a^{-1})\psi(p)-(q+a^{-1})\psi(q)$ |
| Generalized Gamma (MaxEnt for fixed $p$) | $\frac{p}{a^d\Gamma(d/p)}x^{d-1}exp\left(-(x/a)^p\right)$ | $(0,\infty)$ | $E\left(ln\left(X\right)\right)=ln\left(a\right)+p^{-1}\psi\left(\frac{d}{p}\right)$ <br> $E\left(X^p\right)=a^p\frac{d}{p}$ | $ln\left(a\right)+p^{-1}+\ln\left(\Gamma\left(\frac{d}{p}\right)\right)+\left(\frac{d}{p}-1\right)\psi\left(\frac{d}{p}\right)$ |
| Laplace | $\frac{1}{2b}\exp\left(-\frac{\lvert x-u\rvert}{b}\right)$ | $(-\infty,\infty)$ | $E\left(\lvert X-u\rvert\right)=b$ | $1+\ln\left(2b\right)$ |
| Logistic | $\frac{\exp\left(-(x-u)/s\right)}{s\left(1+\exp\left(-(x-u)/s\right)\right)^2}$ | $(-\infty,\infty)$ | $E\left(X\right)=u$ <br> $E\left[\ln\left(1+\exp\left(-\left(x-u\right)/s\right)\right)\right]=\ln\left(2\right)$ | $\ln(s)+2$ |
| Lognormal | $\frac{1}{\sigma x\sqrt{2\pi}}\exp\left(\frac{-\left(\ln\left(x\right)-u\right)^2}{2\sigma^2}\right)$ | $(0,\infty)$ | $E\left(\ln\left(X\right)\right)=u$ <br> $E\left(\left[\ln\left(X\right)\right]^2\right)=u^2+\sigma^2$ | $u+\frac{1}{2}\ln\left(2\pi e\sigma^2\right)$ |
| Normal | $\frac{1}{\sqrt{2\pi\sigma^2}}\exp\left(-\frac{\left(x-u\right)^2}{\sigma^2}\right)$ | $(-\infty,\infty)$ | $E\left(X\right)=u$ <br> $E\left(X^2\right)=u^2+\sigma^2$ | $\frac{1}{2}\ln\left(2\pi e\sigma^2\right)$ |
| Pareto | $\frac{\alpha x_m^\alpha}{x^{\alpha+1}}$ | $(x_m,\infty)$ | $E\left(\ln\left(X\right)\right)=\frac{1}{\alpha}+\ln\left(x_m\right)$ | $\ln\left(\frac{x_m}{\alpha}\right)+1+\frac{1}{\alpha}$ |
| Rayleigh | $\frac{x}{\sigma^2}\exp\left(-\frac{x^2}{2\sigma^2}\right)$ | $(0,\infty)$ | $E\left(X^2\right)=2\sigma^2$ <br> $E\left(\ln\left(X\right)\right)=\frac{\ln\left(2\sigma^2\right)-\gamma_E}{2}$ | $1+\frac{\ln\left(\sigma\right)}{\sqrt{2}}+\frac{\gamma_E}{2}$ |
| Weibull | $\frac{k}{\lambda^k}x^{k-1}\exp\left(-\left(\frac{x}{\lambda}\right)^k\right)$ | $(0,\infty)$ | $E\left(X^k\right)=\lambda^k$ <br> $E\left(\ln\left(X\right)\right)=\ln\left(\lambda\right)-\frac{\gamma_E}{k}$ | $\frac{k-1}{k}\gamma_E+\ln\left(\frac{\lambda}{k}\right)+1$ |

**Note:** $\gamma_E=.5772156649\ldots$ is Euler's constant, $\psi\left(\bullet\right)$ is the digamma function, $\Gamma\left(\alpha\right)$ is the gamma function, and $B\left(\alpha,\beta\right)$ is the Beta function.

## Appendix B. Theorems

### Theorem B.1: Bias in Parametric MLE of Differential Entropy

The bias in the MLE of differential entropy $DE_{ML}\left(f\right)=-\int_{x} f\left(x;\hat{\boldsymbol{\theta}}_{ML}\right)\ln\left(f\left(x;\hat{\boldsymbol{\theta}}_{ML}\right)\right)dx$ is represented by

$$Bias_{MLE}(n)=\frac{1}{2n}tr\left(I^{-1}\left(\boldsymbol{\theta}_0\right)\nabla^2_{\boldsymbol{\theta}}H\left(\boldsymbol{\theta}_0\right)\right)+o\left(n^{-1}\right). \tag{B.1.1}$$

***Proof:*** Define the true value of differential entropy as $DE\left(f\right)=-\int_{x} f\left(x;\boldsymbol{\theta}_0\right)\ln\left(f\left(x;\boldsymbol{\theta}_0\right)\right)dx$. The second order Taylor series expansion of $DE_{ML}\left(f\right)$ around $\boldsymbol{\theta}_0$ under standard ML regularity conditions results in (see Theorem B.2 for justification of $o_p\left(1\right)$):

$$DE_{ML}\left(f\right)\approx DE\left(f\right)+\left(\hat{\boldsymbol{\theta}}_{ML}-\boldsymbol{\theta}_0\right)'\frac{\partial DE\left(f\right)}{\partial\boldsymbol{\theta}_0}+\frac{1}{2}\left(\hat{\boldsymbol{\theta}}_{ML}-\boldsymbol{\theta}_0\right)'\frac{\partial^2 DE\left(f\right)}{\partial\boldsymbol{\theta}_0\partial\boldsymbol{\theta}_0'}\left(\hat{\boldsymbol{\theta}}_{ML}-\boldsymbol{\theta}_0\right)+o_p\left(1\right) \tag{B.1.2}$$

Taking expectations yields

$$Bias_{MLE}\left(n\right)=E\left(\left(\hat{\boldsymbol{\theta}}_{ML}-\boldsymbol{\theta}_0\right)'\frac{\partial DE\left(f\right)}{\partial\boldsymbol{\theta}_0}\right)+\frac{1}{2}E\left(\left(\hat{\boldsymbol{\theta}}_{ML}-\boldsymbol{\theta}_0\right)'\frac{\partial^2 DE\left(f\right)}{\partial\boldsymbol{\theta}_0\partial\boldsymbol{\theta}_0'}\left(\hat{\boldsymbol{\theta}}_{ML}-\boldsymbol{\theta}_0\right)\right)+o\left(n^{-1}\right) \tag{B.1.3}$$

It is known under standard ML regularity conditions that $E\left(\hat{\boldsymbol{\theta}}_{ML}\right)=\boldsymbol{\theta}_0+o\left(n^{-1}\right)$, so that

$$E\left(\left(\hat{\boldsymbol{\theta}}_{ML}-\boldsymbol{\theta}_0\right)'\frac{\partial DE\left(f\right)}{\partial\boldsymbol{\theta}_0}\right)=o\left(n^{-1}\right) \tag{B.1.4}$$

which then implies that the leading bias term is derived from the second order term in the expansion, which can be written as

$$\begin{aligned}\frac{1}{2}E\left(\left(\hat{\boldsymbol{\theta}}_{ML}-\boldsymbol{\theta}_0\right)'\frac{\partial^2 DE\left(f\right)}{\partial\boldsymbol{\theta}_0\partial\boldsymbol{\theta}_0'}\left(\hat{\boldsymbol{\theta}}_{ML}-\boldsymbol{\theta}_0\right)\right)&=\frac{1}{2}tr\left(E\left(\left(\hat{\boldsymbol{\theta}}_{ML}-\boldsymbol{\theta}_0\right)\left(\hat{\boldsymbol{\theta}}_{ML}-\boldsymbol{\theta}_0\right)'\right)\frac{\partial^2 DE\left(f\right)}{\partial\boldsymbol{\theta}_0\partial\boldsymbol{\theta}_0'}\right)\\&=\frac{1}{2}tr\left(Cov\left(\hat{\boldsymbol{\theta}}_{ML}\right)\frac{\partial^2 DE\left(f\right)}{\partial\boldsymbol{\theta}_0\partial\boldsymbol{\theta}_0'}\right)\\&=\frac{1}{2n}tr\left(I^{-1}\left(\boldsymbol{\theta}_0\right)\frac{\partial^2 DE\left(f\right)}{\partial\boldsymbol{\theta}_0\partial\boldsymbol{\theta}_0'}\right)\end{aligned}$$

where $I^{-1}(\boldsymbol{\theta}_0)$ is the inverse of Fisher's information matrix, $Cov(\hat{\boldsymbol{\theta}}_{ML}) = n^{-1}I^{-1}(\boldsymbol{\theta}_0)$, and

$$\nabla_{\boldsymbol{\theta}}^2 H(\boldsymbol{\theta}_0) = \frac{\partial^2 DE(f)}{\partial \boldsymbol{\theta}_0 \partial \boldsymbol{\theta}_0'}. \qquad \square$$

**Theorem B.2: Derivation of the KDE Differential Entropy Representation in Ln-Space**

Let $X \sim f(x)$ and $Y = \ln(X) \sim g(y)$. Then $DE(f) = DE(g) + E_g(Y)$.

***Proof:*** By a change of variables, $x = e^y \Rightarrow dx = e^y dy$, it follows that

$$\begin{aligned} DE(f) &= -\int_0^\infty f(x)\ln(f(x))dx = -\int_{-\infty}^\infty f(e^y)\ln(f(e^y))e^y dy \\ &= -\int_{-\infty}^\infty g(y)\ln(e^{-y}g(y))dy = -\int_{-\infty}^\infty g(y)\left[\ln(g(y)) - y\right]dy \\ &= DE(g) + E_g(Y) \qquad \square \end{aligned}$$

**Theorem B.3. Joint Asymptotic Normality of $\hat{\boldsymbol{\theta}}_n$ and $DE_{KDE}(\hat{f};h)$**

Let $\{X_1, X_2, \cdots, X_n\} \sim iid\ f(x;\boldsymbol{\theta}_0)$ be a random sample from a parametric family of densities indexed by $\boldsymbol{\theta} \in \Theta \subset \mathbb{R}^k$, and $\boldsymbol{\theta}_0 \in \text{int}(\Theta)$ is the true parameter value. Let $\hat{\boldsymbol{\theta}}_{ML}$ be the MLE of $\boldsymbol{\theta}_0$ based on the random sample and let $DE_{KDE}(\hat{f};h)$ represent the KDE of differential entropy. Assume the following conditions hold:

i. $f(x;\boldsymbol{\theta}) \in C^2$ in $\boldsymbol{\theta}$, and is identifiable at $\boldsymbol{\theta}_0$.
ii. Classical ML regularity conditions hold (van der Vaart (1998), Theorem 5.23), including allowing differentiation w.r.t. $\boldsymbol{\theta}$ under integral signs for integrands relating to the expected score function, Fisher information, and higher order derivatives of the log-likelihood function, and

   a. $\nabla_{\boldsymbol{\theta}} \ln(f(x;\boldsymbol{\theta}))$ exists and has finite expectation and variance,
   b. $\hat{\boldsymbol{\theta}}_{ML}$ is the solution to the score function $n^{-1}\sum_{i=1}^n \nabla_{\boldsymbol{\theta}} \ln(f(x_i;\boldsymbol{\theta})) = \mathbf{0}$,
   c. $\sqrt{n}(\hat{\boldsymbol{\theta}}_{ML} - \boldsymbol{\theta}_0) \xrightarrow{d} \text{N}(\mathbf{0}, \mathbf{I}(\boldsymbol{\theta}_0)^{-1})$ where $\mathbf{I}(\boldsymbol{\theta}_0)$ denotes the positive definite Fisher information matrix

$$E_{\boldsymbol{\theta}_0}\left[\nabla_{\boldsymbol{\theta}_0} \ln(f(X;\boldsymbol{\theta}_0)) \nabla_{\boldsymbol{\theta}_0} \ln(f(X;\boldsymbol{\theta}_0))'\right],$$

iii. The KDE $DE_{KDE}\left(\hat{f};h\right)$ is based on a smooth and strictly positive valued kernel (e.g., Gaussian), with a regular bandwidth for which $h_n \to 0$ and $nh_n \to \infty$.

Then

$$n^{1/2}\begin{bmatrix}\left(DE_{KDE}\left(\hat{f};h\right)-DE(f\left(x;\boldsymbol{\theta}_0\right))\right)\\ \left(\hat{\boldsymbol{\theta}}_n-\boldsymbol{\theta}_0\right)\end{bmatrix}\xrightarrow{d}N\left(\mathbf{0},\begin{bmatrix}\sigma^2_{DE_{KDE}} & \mathbf{cov}\left(DE_{KDE},\hat{\boldsymbol{\theta}}\right)'\\ \mathbf{cov}\left(DE_{KDE},\hat{\boldsymbol{\theta}}\right) & \mathbf{I}^{-1}(\boldsymbol{\theta}_0)\end{bmatrix}\right).$$

***Proof:*** A centered and scaled version of the ML estimator of $\boldsymbol{\theta}$ can be represented via an influence function representation as

$$\sqrt{n}\left(\hat{\boldsymbol{\theta}}_{ML}-\boldsymbol{\theta}_0\right)=n^{-1/2}\sum_{i=1}^{n}\boldsymbol{\psi}_{ML}(X_i)+o_p(1) \tag{B.3.1}$$

where the classical influence function for the MLE is defined as $\boldsymbol{\psi}_{ML}(x)=\mathbf{I}(\boldsymbol{\theta}_0)^{-1}\nabla_{\boldsymbol{\theta}}\ln\left(f\left(x;\boldsymbol{\theta}_0\right)\right)$, with $\mathrm{E}\left(\boldsymbol{\psi}_{ML}(X)\right)=\mathbf{0}$ and $\mathbf{Cov}\left(\boldsymbol{\psi}_{ML}(X)\right)=\mathbf{I}(\boldsymbol{\theta}_0)^{-1}$ (See van der Vaart (1998), Theorem 5.23 and Example 5.34.)

Similarly, a centered and scaled version of the KDE of differential entropy can be represented as

$$\sqrt{n}\left(DE_{KDE}\left(\hat{f};h\right)-DE(f\left(x;\boldsymbol{\theta}_0\right))\right)=\ n^{-1/2}\sum_{i=1}^{n}\phi_{KDE}(X_i)+o_p(1) \tag{B.3.2}$$

where the KDE influence function is defined as $\phi_{KDE}(x)=-\ln\left(f\left(x;\boldsymbol{\theta}\right)\right)-DE(f\left(x;\boldsymbol{\theta}_0\right)$, with $E\left(\phi_{KDE}(X)\right)=0$ and $\mathrm{var}\left(\phi_{KDE}(X)\right)=\sigma^2_{DE_{KDE}}$ (See Joe (1989) and Hall and Morton (1993)).

Then from (B.1.1) and (B.1.2) the following joint representation of the MLE of the parameter vector and KDE estimator of differential entropy can be defined as

$$\sqrt{n}\begin{pmatrix}DE_{KDE}\left(\hat{f};h\right)-DE(f\left(x;\boldsymbol{\theta}_0\right)\\ \hat{\boldsymbol{\theta}}_n-\boldsymbol{\theta}_0\end{pmatrix}=\frac{1}{\sqrt{n}}\sum_{i=1}^{n}\begin{pmatrix}\phi_H(X_i)\\ \boldsymbol{\psi}(X_i)\end{pmatrix}+\boldsymbol{o}_p(1)=\frac{1}{\sqrt{n}}\sum_{i=1}^{n}\mathbf{Z}_i+\boldsymbol{o}_p(1). \tag{B.3.3}$$

where $\mathbf{Z}_i = \begin{bmatrix} \phi_H(X_i) \\ \boldsymbol{\psi}(X_i) \end{bmatrix}$. The since the $\mathbf{Z}_i$'s are independent and identically distributed and given that $\mathbf{Cov}(\mathbf{Z}) = \begin{bmatrix} \sigma^2_{DE_{KDE}} & \mathbf{cov}\left(DE_{KDE}, \hat{\boldsymbol{\theta}}_{ML}\right)' \\ \mathbf{cov}\left(DE_{KDE}, \hat{\boldsymbol{\theta}}_{ML}\right) & \mathbf{I}^{-1}(\boldsymbol{\theta}_0) \end{bmatrix}$, the Multivariate Lindberg-Levy CLT (van der Vaart (1998), Theorem 2.27) can be applied to support the conclusion of the theorem. □

**Theorem B.4: Asymptotic Normality of DDE**

Let $\{X_1, X_2, \cdots, X_n\} \sim iid\ f(x;\boldsymbol{\theta}_0)$ be a random sample from a parametric family of densities indexed by $\boldsymbol{\theta} \in \Theta \subset \mathbb{R}^k$, and $\boldsymbol{\theta}_0 \in \text{int}(\Theta)$ is the true parameter value. Let $\hat{\boldsymbol{\theta}}_{ML}$ be the MLE of $\boldsymbol{\theta}_0$ based on the random sample and $DE_{KDE}\left(\hat{f};h\right)$ represent the KDE of differential entropy. Assume the conditions of Theorem B.3 hold. Then,

$$n^{1/2}DDE(f) = n^{1/2}\left(DE_{ML}(f(x;\boldsymbol{\theta})) - DE_{KDE}\left(\hat{f};h\right)\right) \xrightarrow{d} N\left(0, \sigma^2_{DDE}\right)$$

where

$$\sigma^2_{DDE} = \sigma^2_{DE_{KDE}} + \nabla_{\boldsymbol{\theta}_0} DE(f(x;\boldsymbol{\theta}_0)' \mathbf{I}^{-1}(\boldsymbol{\theta}_0) \nabla_{\boldsymbol{\theta}_0} DE(f(x;\boldsymbol{\theta}_0) - 2\nabla_{\boldsymbol{\theta}_0} DE(f(x;\boldsymbol{\theta}_0)' \mathbf{cov}\left(DE_{KDE}, \hat{\boldsymbol{\theta}}_{ML}\right).$$

***Proof:*** Consider the mean value form of the Taylor series expansion of differential entropy evaluated at the maximum likelihood estimator as

$$DE\left(f\left(x;\hat{\boldsymbol{\theta}}_{ML}\right)\right) = DE(f(x;\boldsymbol{\theta}_0) + \nabla_{\boldsymbol{\theta}} DE(f(x;\boldsymbol{\theta})\big|_{\boldsymbol{\theta}_0}' \left(\hat{\boldsymbol{\theta}}_{ML} - \boldsymbol{\theta}_0\right) + R_n,$$

where the remainder term $R_n = \frac{1}{2}\left(\hat{\boldsymbol{\theta}}_n - \boldsymbol{\theta}_0\right)' \nabla^2_{\boldsymbol{\theta}} DE(f(x;\boldsymbol{\theta}))\big|_{\boldsymbol{\theta}_*} \left(\hat{\boldsymbol{\theta}}_{ML} - \boldsymbol{\theta}_0\right)$, for some $\boldsymbol{\theta}_*$ between $\hat{\boldsymbol{\theta}}_{ML}$ and $\boldsymbol{\theta}_0$, satisfies $R_n = o_p\left(n^{-1/2}\right)$. Then the DDE statistic can be written as

$$\begin{aligned} DDE(f) &= \left(DE\left(f\left(x;\hat{\boldsymbol{\theta}}_{ML}\right)\right) - DE_{KDE}\left(\hat{f};h\right)\right) \\ &= \nabla_{\boldsymbol{\theta}} DE(f(x;\boldsymbol{\theta})\big|_{\boldsymbol{\theta}_0} \left(\hat{\boldsymbol{\theta}}_{ML} - \boldsymbol{\theta}_0\right) - \left(DE_{KDE}\left(\hat{f}\right) - DE(f(x;\boldsymbol{\theta}_0)\right) + R_n \end{aligned}$$

It follows that

$$n^{1/2}DDE(f) = \nabla_{\boldsymbol{\theta}} DE(f(x;\boldsymbol{\theta}))\Big|_{\boldsymbol{\theta}_0}' \, n^{1/2}\left(\hat{\boldsymbol{\theta}}_{ML} - \boldsymbol{\theta}_0\right) - n^{1/2}\left(DE_{KDE}\left(\hat{f}\right) - DE^f(\boldsymbol{\theta}_0)\right) + o_p(1)$$

which to order $o_p(1)$ represents a linear transformation of the multivariate Normal distribution stated in the conclusion of theorem B.3., which then justifies the conclusions of the theorem. □

**Theorem B.5: Consistency of the DDE Test**

Under the conditions stated in Theorem B.2, the test of distributional hypotheses based on the $DDE(f)$ statistic will be consistent against fixed alternatives not encompassed by $H_0$.

***Proof:*** Let distribution $\xi$ be such that $x_i \sim iid\ \xi$ where $\xi \neq f\ \forall \boldsymbol{\theta}$, and define

$$DDE(f) = \left(DE_{ML}(f) - DE(\xi)\right) + \left(DE(\xi) - DE_{KDE}\left(\hat{f}\right)\right)$$

where $DE(\xi)$ denotes the differential entropy associated with the distribution $\xi(x)$. Under the conditions of Theorem B.1,

- $DE_{KDE}\left(\hat{f}\right) \xrightarrow{p} DE(\xi)$
- $\hat{\boldsymbol{\theta}}_{ML} \xrightarrow{p} \boldsymbol{\theta}^* = \arg\min_{\boldsymbol{\theta}} \left\{ D_{KL}\left(\xi \| f(x;\boldsymbol{\theta})\right) \right\}$, where $D_{KL}$ denotes Kulback-Leibler divergence (note, this is the standard MLE convergence result under model misspecification)
- $DE_{ML}(f) \xrightarrow{p} DE\left(f\left(x;\boldsymbol{\theta}^*\right)\right)$

It follows that

$$DDE(f) \xrightarrow{p} \left(DE\left(f\left(x;\boldsymbol{\theta}^*\right)\right) - DE(\xi)\right) = \tau \neq 0$$

so that

$$n^{1/2}\left|DDE(f)\right| \xrightarrow{p} n^{1/2}\left|\left(DE\left(f\left(x;\boldsymbol{\theta}^*\right)\right) - DE(\xi)\right)\right| = n^{1/2}|\tau| \to \infty$$

and $$\lim_{n\to\infty} P\left(\left|DDE(f)\right| > \delta\right) = 1\ \forall \delta < \infty. \quad □$$

## Appendix C. Rationale for Bandwidth Definitions

This section provides a unified description of the automated bandwidth-selection rule used for the KDE in computing the entropy-gap statistic. The rule is designed to stabilize Type-I error rates under the parametric null while maintaining monotonically increasing power as sample size increases across a wide range of non-Gaussian alternatives. It synthesizes smoothness-based guidance from the asymptotic literature on differential-entropy estimation and bandwidth selection (e.g., Joe 1989; Beirlant et al. 1997; Moon et al. 1995; Sricharan et al.; Delyon and Portier 2016) with simulation-based evidence for the entropy-gap test experienced here. The section also provides a description of the small sample inflation factor $k(n)$.

The central tuning parameter is the multiplicative constant $c$ in the bandwidth expression $h = k(n)c\hat{\sigma}n^{-1/5}$, where $\hat{\sigma}$ is the sample standard deviation of either the raw data for $X \in \mathbb{R}$ or *ln*-transformed data for $X \in \mathbb{R}_+$, and $n$ is the sample size. The factor $c$ affects the degree of smoothing of the KDE estimate and directly influences the small-sample bias of the nonparametric entropy estimator. The selection of $c$ follows a regime-structured rule depending on the support of the null model, its smoothness, and its tail behavior.

### *C.1. Gaussian and Near-Gaussian Null Distributions*

If the fitted parametric null distribution is supported on the real line and is approximately symmetric and Gaussian-like, then the optimal bias–variance tradeoff for entropy estimation is typically achieved at $c = 1$. This recommendation is consistent with classical bandwidth theory for smooth, symmetric densities (e.g., Silverman 1986; Wand and Jones 1995).

To operationalize the concept of "near-Gaussian," we implement the following three identifying conditions:

1. Sample is supported on the real line: $X \in \mathbb{R}$.

2. Approximate symmetry, as indicated by the sample skewness measure

$$|\hat{s}_{skew}| = \left| \frac{\frac{1}{n}\sum_{i=1}^{n}(X_i - \bar{X})^3}{\left(\frac{1}{n}\sum_{i=1}^{n}(X_i - \bar{X})^2\right)^{3/2}} \right| \leq .5 .$$

3. Moderate kurtosis, as categorized by the sample kurtosis

$$\hat{\kappa} = \frac{\frac{1}{n}\sum_{i=1}^{n}(X_i - \bar{X})^4}{\left(\frac{1}{n}\sum_{i=1}^{n}(X_i - \bar{X})^2\right)^2} \in [2,4].$$

When all three conditions are satisfied, the value of $c$ is set to 1.

*C.2. Non-Gaussian Nulls With Real-Line Support*

If the null distribution is supported on the real line but is not Gaussian-like (e.g., Laplace, logistic with sharp tails, symmetric hyperbolic distributions), the KDE bias differs markedly from the Gaussian case. Sharp peaks and heavier tails imply that modest undersmoothing improves entropy estimation accuracy, consistent with recommendations in Joe (1989), Beirlant et al. (1997), and Delyon and Portier (2016).

For these distributions, a kurtosis-adaptive rule: $c = 1 + 0.1\log_2(\hat{\gamma}_{\text{kurt}})$ is implemented, where $\hat{\gamma}_{\text{kurt}} = \frac{\kappa_0}{\tau(\hat{\kappa})}$, $\kappa_0$ is the kurtosis implied by the fitted parametric null distribution, $\tau(\hat{\kappa}) = \min\{\max(\hat{\kappa}, 2), 10\}$, $\hat{\kappa}$ is the sample kurtosis as defined above, and $c$ is constrained to the interval $[0.85, 1.15]$ for stability. This rule shrinks $c$ below 1 for heavy-tailed or sharp-peaked densities and modestly increases $c$ for smooth lighter-tailed real-line densities.

This structure is motivated by both theoretical and methodological reasons. The entropy-gap statistic compares a generally biased parametric entropy estimator, $\hat{h}_{\text{ML,null}}$, with a biased nonparametric log-KDE entropy estimator, $\hat{h}_{\text{KDE}}$. Because the test is calibrated through a parametric bootstrap under the null distribution, valid Type I error control depends on ensuring that the bias of $\hat{h}_{\text{KDE}}$ in the observed data matches the bias of the same estimator across bootstrap replicates. Since the bootstrap samples are drawn from the null model itself, their smoothness, curvature, and tail behavior are reflected in the null distribution's kurtosis $\kappa_0$. Leading bias terms in kernel-based entropy estimation depend on these global shape features (Joe 1989; Moon et al. 1995; Delyon and Portier 2016). For these reasons the bandwidth used for the KDE, for both the original data sample and bootstrap sampling, is anchored to the same smoothness class, namely, the one represented by $\kappa_0$.

The term $\tau(\hat{\kappa})$ stabilizes the sample kurtosis by truncating extreme values that can occur due to finite-sample variability. Its ratio $\frac{\kappa_0}{\tau(\hat{\kappa})}$ measures whether the empirical sample exhibits heavier or lighter tails, or sharper or smoother curvature, than expected under the null. When $\hat{\kappa} \approx \kappa_0$, the ratio is near one and $c \approx 1$, yielding the same general level of smoothing in the observed and bootstrap samples. If the sample density appears heavier-tailed (larger $\hat{\kappa}$), the ratio decreases and $c$ is decreased, providing modest undersmoothing to counteract larger curvature. Conversely, if the empirical data appear smoother than the null (smaller $\hat{\kappa}$), the ratio exceeds one and $c$ is increased, resulting in modest oversmoothing. The adjustments are centered on $\kappa_0$ so that the null-model bias profile is preserved across real and bootstrap samples.

Note that using any fixed "reference" kurtosis in place of $\kappa_0$ would disrupt alignment. The observed sample and the bootstrap samples would utilize different effective smoothing, the bias structure of $\hat{h}_{\text{KDE}}$ would diverge between the two sampling contexts, and the bootstrap approximation to the null distribution of the entropy-gap statistic would be misrepresented. Consequently, anchoring the adaptive bandwidth to the null distribution kurtosis is relevant for ensuring stability, Type I error control, and addressing potential bias propagation within the entropy-gap testing framework.

*C.3. Non-Gaussian Nulls With Positive Support*

If the null distribution is supported on $(0,\infty)$ and is unimodal and right-skewed (e.g., Gamma, Lognormal, inverse Gaussian, Weibull, Generalized Beta), the *ln*-transform is applied to the data before the KDE is implemented. The transformation produces an approximate real-line support that addressed boundary (at zero) issues but does not remove asymmetry or tail-thickness effects in entropy estimation. In such settings, KDE bias is strongly determined by the tail behavior of $ln(X)$; see Moon et al. (1995), Sricharan et al. (2012), and Delyon and Portier (2016). Simulations confirm that using $c=1$ can lead to substantial under- or over-smoothing and unstable Type-I error and power calculations when the null distribution is not Gaussian-like. For all such positive-support right-skewed null distributions, we apply the same kurtosis-adaptive rule as was described in section C.2, and for the same reasons.

### *C.4. Small Sample Inflation Factor*

Simulation studies (Joe, 1989; Moon et al., 1995) reveal that at small $n$, KDE-based entropy estimators tend to under-smooth, leading to conservative Type I error rates. To mitigate this, the bandwidth is inflated by a sample-size-dependent multiplier $k(n)$ defined by

$$k(n) = \begin{Bmatrix} 1.25 - 0.25\left(\dfrac{n-50}{50}\right) \\ 1 \end{Bmatrix} \; for \; \begin{Bmatrix} n < 100 \\ n \geq 100 \end{Bmatrix}$$

so that for example, $k(50) = 1.25$ and $k(100) = 1.00$. The inflation of the bandwidth compensates for finite-sample bias in entropy estimation similar to the variance-correcting factors suggested by Hall and Marron (1987) and used by Singh et al. (2003) in small-sample KDE functional estimation. Beyond $n = 100$, the asymptotic calibration dominates, and $k(n) = 1$ is retained.

### *C.5. Summary of the Bandwidth Rule*

Let $\hat{\kappa}$ and $\hat{s}_{skew}$ denote the sample kurtosis and skewness of the raw data, and let $\kappa_0$ denote the kurtosis implied by the fitted parametric null distribution. The bandwidth is defined as

$$h = k(n)c\hat{\sigma}n^{-1/5} \quad \text{where } c = \begin{Bmatrix} 1 \\ 1 + 0.1\log_2\left(\dfrac{\kappa_0}{\tau(\hat{\kappa})}\right) \end{Bmatrix} if \begin{Bmatrix} X \in \mathbb{R}, \hat{s}_{skew} \leq .5, \; \hat{\kappa} \in [2,4] \\ otherwise \end{Bmatrix}$$

with truncation $\tau(\hat{\kappa}) = \min\{\max(\hat{\kappa},2),10\}$ *and* $c \in [0.85, 1.15]$, and both $\hat{\sigma}$ and the KDE are calculated from either the raw or *ln*-transformed data depending on whether $X \in \mathbb{R}$ *or* $\mathbb{R}_+$. This rule supports optimal smoothing for Gaussian and Gaussian-like nulls, as well as bias-adaptive smoothing for both non-Gaussian null distributions on $\mathbb{R}$ and skewed null distribution on $\mathbb{R}_+$. Within the adaptive bandwidth framework heavier-tailed or sharper-peaked densities are subject to modest undersmoothing, while smoother, lighter-tailed distributions receive modest oversmoothing.

All recommendations are grounded in entropy-estimation theory and KDE bias expansions (Joe 1989; Moon et al. 1995; Sricharan et al.; Beirlant et al. 1997; Delyon and Portier 2016) and are also supported by Monte Carlo experiments.